\documentclass{article}
\usepackage{float}
\usepackage{graphicx}
\usepackage{amsmath}
\usepackage{caption}
\usepackage{subcaption}
\usepackage[margin=0.5in]{geometry}
\usepackage{lineno}

\usepackage{algorithm}
\usepackage{algorithmic}

\usepackage{siunitx}
\usepackage{threeparttable}

\title{Capturing Finite Target Dynamics: Phase-Delayed Analytic Modeling of Multi-Layer Penetration Events}
\author{Trenton Kirchdoerfer\thanks{Lawrence Livermore National Laboratory Email: kirchdoerfer1@llnl.gov}}

\begin{document}
\maketitle

\begin{abstract}

    \centering
    \fbox{
    \parbox{0.7\textwidth}{

    The Walker-Anderson half-space penetration model has long been successfully used for the rapid, efficient calculation of penetration depth by rigid and eroding rods.
    These models align well with detailed simulations for thick targets; however, existing extensions for finite targets struggle to accurately capture nose-tail velocity profiles in thinner targets.
    For stack-ups of thin-walled targets, this deficiency results in mischaracterized rod-erosion relative to hydro-simulations or experimental predictions.
    In this work, we leverage insights from detailed hydro-simulations to propose an updated modification to the Walker-Anderson model to correctly account for wave propagation within a given target.
    This addition improves results for thin targets while retaining good behavior for thick targets with zero additional model parameters.
    Our updated model exhibits strong agreement with detailed simulations for penetrations through multiple thin targets.
     }}

\end{abstract}

\section{Introduction}

Within penetration mechanics, analytic models have a significant history of providing the ability to both summarize and efficiently predict effects within targets and projectiles \cite{Anderson2017}.
Such capabilities also exist in modern general-simulation hydro-dynamic codes, but at a serious increase in code complexity, numerical expense and required operator expertise.
For instance, our longest duration simulation result in Section \ref{sec:Results} required a compiled hydro-code to run for 11 minutes on a 112 processor compute node to achieve comparable results to an interpreted language implementation of our analytic model operating on a single processor of a laptop.
Additionally, those general simulation codes frequently require fine-tuning of stability settings to keep the millions of degrees of freedom stable during a set of extremely dynamic processes.
This work expands on methods \cite{Walker1995} that do require numerical time integration, however these methods offer stability benefits and massive complexity reductions, making these models highly useful for engineers developing tools for fast reliable answers.

We specifically note that our extensions to such models is at odds with the significant trends in fast-solving models for penetration mechanics that invoke machine learning strategies \cite{Anderson2015fitting,Rietkerk2023,Ryan2024} that are themselves children of traditional curve/trend fits \--- albeit with greater potential to match complex behavior.
However semi-analytic, mechanics-informed methods have not been reduced in their efficacy by time and can additionally provide intuition that generalized fits cannot easily supply.
  
For eroding rod projectiles, Walker and Anderson \cite{Walker1995} developed a robust framework of assumed flow field models built from intuitions they derived from detailed 2D hydro-simulations for half-space impacts.
This model was later augmented by a number of extensions to address finite domains \cite{Ravid1998,Chocron2003,Walker1999bulge}, alternate material types \cite{Walker1998ceramic,Zolfaghari2017concrete,Bavdekar2019ceramic}, and even composite targets \cite{Walker2006layer}.
Within these families of extensions, this paper will seek to improve the modeling of finite monolithic targets, specifically focusing on thinner domains than was the focus of previous works.
  
\begin{figure}[H]
    \centering
    \fbox{\includegraphics[width=0.5\textwidth]{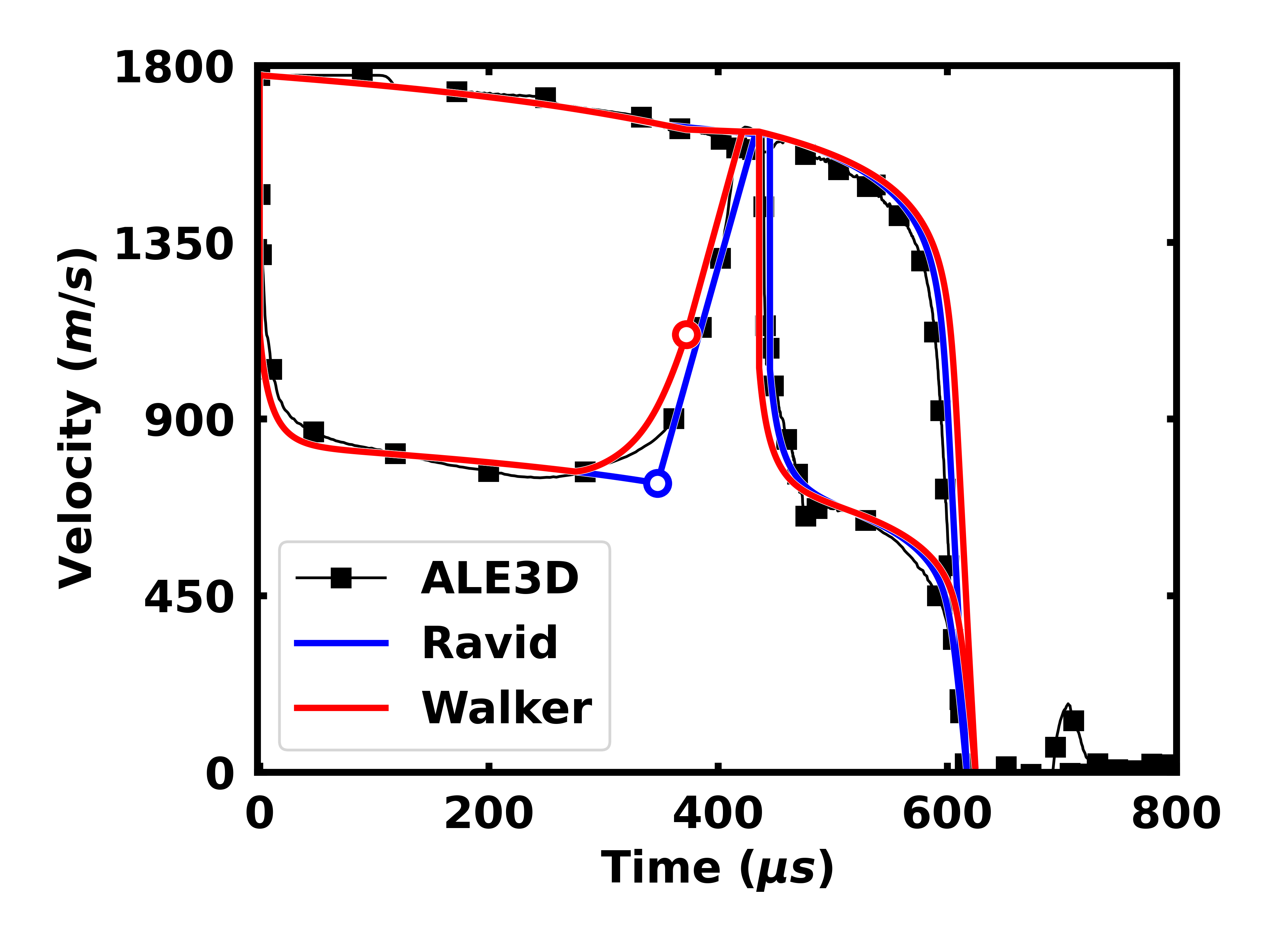}}
    \caption{Previous models compared against hydro-simulation result, including both projectile nose and tail velocities; hollow circles indicate target failure. Lower and higher of a given style respectively represent nose and tail response of the projectile.  }
    \label{fig:thicktest}
\end{figure}

Principally, the efforts that updated the half-space to a finite domain act as corrections on the original infinite-depth model.
The first model developed by Ravid \cite{Ravid1998} introduced a geometric bulge formation and failure criteria without modifying the original half-space balance equations.
There failure was assessed with time integrated strain-criteria which estimated the time at which the target provided no additional resistance.
It was followed \cite{Chocron2003} with an extension to handle projectile recovery after target failure and was applied to multi-target impacts which inspire our own demonstrations later in the paper.
A separate back-face model was also developed by Walker \cite{Walker1999bulge}, which also included bulging and failure \cite{Chocron1999} within the target but with a updated momentum balance that properly addressed the finite domain.
Both models are remarkably effective at capturing finite target response, and a comparison of their performance against a hydro-simulation as shown in Figure~\ref{fig:thicktest}.
This borrowed \cite{Chocron2003} scenario is made up of a 49.9 cm tungsten projectile of 1.65 cm diameter as it impacts a 30 cm thick steel target followed by a half-space witness plate 10 cm behind the finite target.
Within the plot, we see the nose and tail velocity profiles of the two discussed models compared against a matched hydro-simulation result.
Note that the plot styles are matching for nose and tail history traces as there is no ambiguity: nose velocity is lower and tail velocity is higher.
The hollow dots indicate when the respective analytic models affect failure.
In Figure~\ref{fig:thicktest}, we see that the Ravid model performs well, and we find it's strain-failure cutoffs are relatively easy to tune for triggering the projectile recovery phase of the calculation. 
But despite those results, it is apparent that the Walker back-face model more correctly captures nose reacceleration via the more rigorous treatment of the momentum balance.
As can be seen by the smooth transition after failure, this treatment results in the projectile recovery phase becoming largely insensitive to the specific failure strains.
Between the two models then, it is apparent that Walker's approach to the finite domain better captures the dynamics of a weakening target.
  
\begin{figure}[H]
    \centering
    \begin{subfigure}[t]{0.48\textwidth}
        \centering
        \fbox{\includegraphics[width=0.98\textwidth]{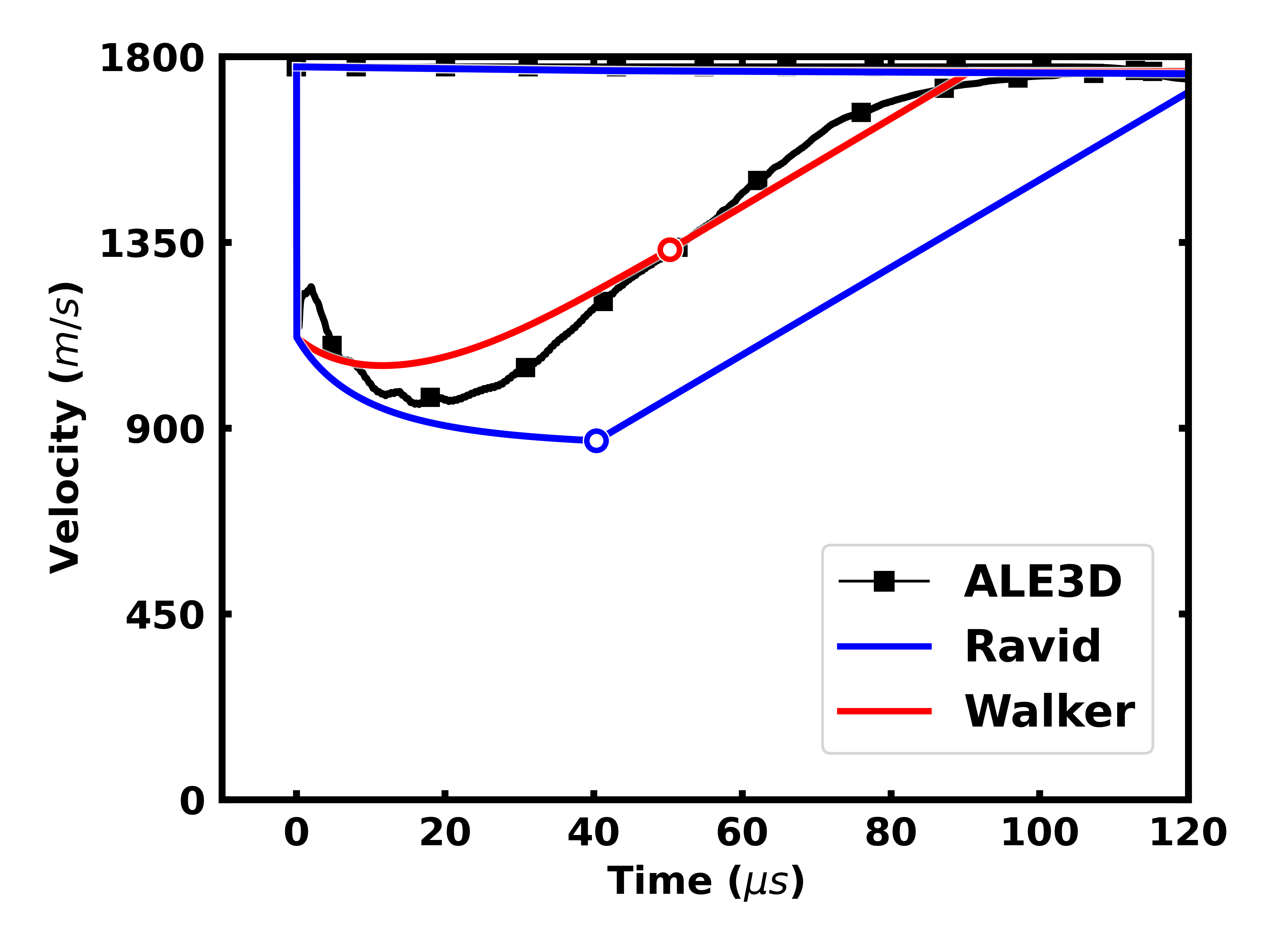}}
    \caption{Nominal diameter projectile (1.65 cm)}
        \label{fig:thintestx1}
    \end{subfigure}
    ~
    \begin{subfigure}[t]{0.48\textwidth}
        \centering
        \fbox{\includegraphics[width=0.98\textwidth]{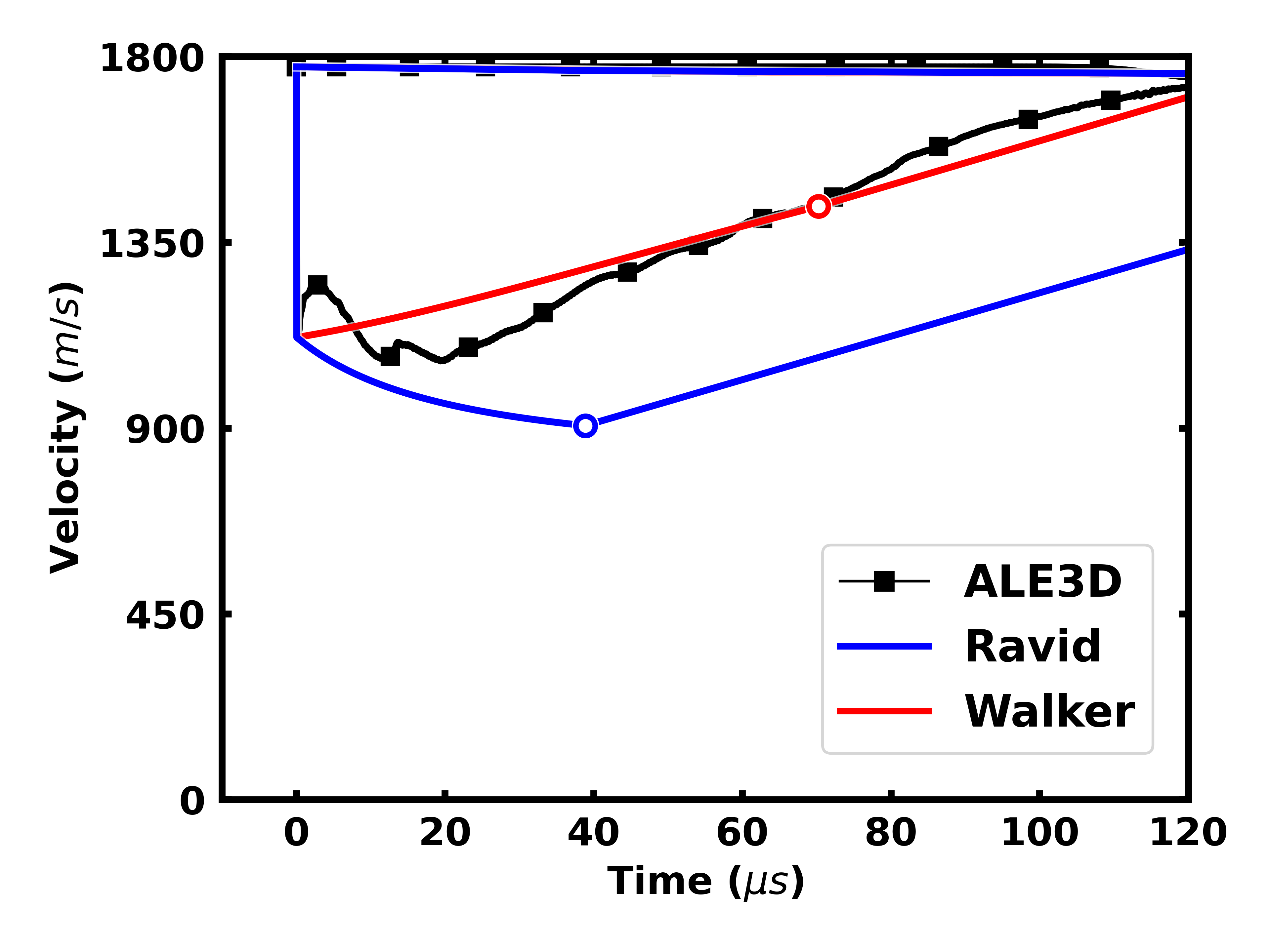}}
    \caption{Double diameter projectile (3.3 cm)}
        \label{fig:thintestx2}
    \end{subfigure}
    \caption{Model nose and tail velocities demonstrated against 4 cm target; hollow circles indicate target failure. }
    \label{fig:thintests}
\end{figure}

However, as we examine impacts into thinner targets, the possibility for improvement becomes more apparent.
To demonstrate, we modify our earlier problem so that the same projectile impacts a single 4 cm target and plot the results in Figure~\ref{fig:thintestx1}.
In general we see that all significant model differences occur in the velocity drop of the projectile nose.
Within that context, we see that prior to 15 $\mathrm{\mu s}$ the Ravid model treatment of the target as infinite yields reasonable results, though after that time the model falters.
Alternatively, we again see the Walker model matches the smooth character of the hydro-simulation projectile recovery, but now it is clear that the initiation of that recovery is premature.
In this model, the Walker finite target effect becomes activated as soon as the plastic zone reaches the target back-face. 
To explore this behavior further, we scale up the projectile diameter, and consequently the plastic zone, which exaggerates the effect shown in Figure~\ref{fig:thintestx2}
What we see in both Figure \ref{fig:thintests} plots is that an infinite target model-assumption is outperforming our more advanced finite target model in early time, when the finite target effects have not yet had time to propagate across the target domain. 
Incorporating that understanding into an updated model will then be the primary focus of this paper, with some effort expended to improve early time response as well.
Given that our reference models are detailed hydro-simulation results, we can interrogate the target response to better understand how best to advance the model. 
Such advancements will be shown to improve our analytic model performance without heavy computational penalties.

The structure of the paper is as follows:
First, we review the original Walker-Anderson \cite{Walker1995} model and identify parameters for both the model and reference simulations, which serve as the basis for subsequent comparisons.
Next, we investigate improvements in impact initialization to enhance the early-time response for infinite targets.
With these updates we then proceed to review the critical aspects of the Walker modified flow model \cite{Walker1999bulge} that affect a finite domain.
We demonstrate that directly applying improved impact initialization to finite target models introduces challenges, highlighting the need to account for wave propagation effects.
By post-processing hydro-simulation results, we examine the influence of non-linear wave propagation and show that linear bulk-wave propagation adequately captures the essential mechanics.
This insight guides the implementation of a phase delay consistent with wave propagation in our model.
To experimentally validate our target modeling assumptions, comparisons are provided against available data on thick finite targets \cite{AndersonStilp1995}.
Finally, we present detailed comparisons across targets of varying thicknesses and velocities, as well as multi-target impacts, using detailed hydro-simulations as reference solutions.
The paper concludes with a summary of the findings and a discussion of remaining model inconsistencies.

\section{The Model} \label{sec:themodel}

\subsection{Review of the Half-space Model}

\begin{figure}[H]
    \centering
    \includegraphics[width=0.4\textwidth]{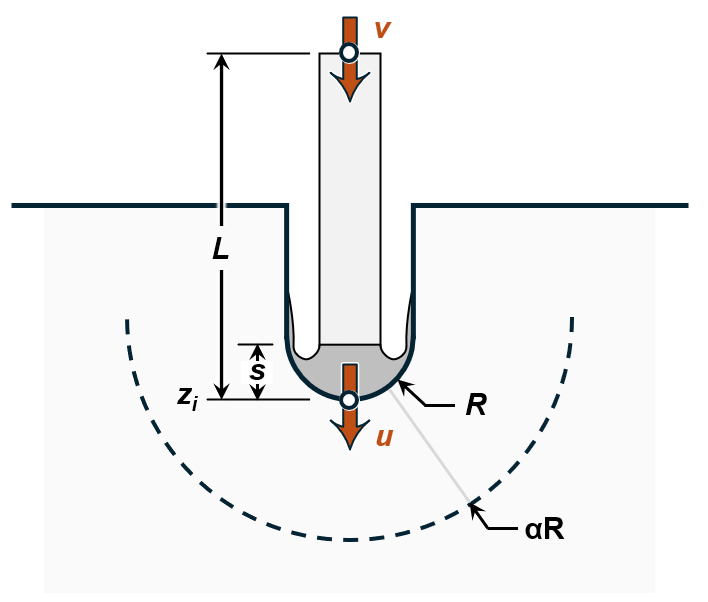}
    \caption{Model layout of projectile penetration in a half-space model}
    \label{fig:infsketch}
\end{figure}

We begin by first reviewing the critical elements of both the original Walker-Anderson \cite{Walker1995} half-space model.
Figure~\ref{fig:infsketch} sketches the geometric components  of the half-space model.
Velocities of the nose and tail are denoted $u$ and $v$, with their corresponding coordinates being $z_i$ and $z_p$,respectively.
The parameter $L$ denotes the total length of the projectile, with its plastic extent represented by $s$.
Specifically within the target, the extent of the assumed flow field within the target is a scale factor, $\alpha$, acting on the crater radius, $R$.

We assume the crater radius is constant, with its calculation being driven by an empirical relation between impact velocity $v_0$, and projectile radius $R_p$, such that
\begin{equation}
        R = R_p (1 + 0.287 v_0 + 0.148 v_0^2).
    \label{eq:cratercalc}
\end{equation}
The momentum balance equation,
\begin{equation}
    \rho_p \dot{v}(L-s)  
    + \dot{u}\left\{\rho_ps + \rho_t R \frac{\alpha - 1}{\alpha + 1}\right\}
    + \rho_p \ \frac{\mathrm{d}}{\mathrm{d}t}\!\!\left( \frac{v-u}{s} \right)\frac{s^2}{2}
    + \rho_t \dot{\alpha}\frac{2Ru}{(\alpha+1)^2} 
    \ \  = \ \  
    \frac12\rho_p(v-u)^2 - \left\{\frac12\rho_t u^2 + \frac73 Y_t \ln{\alpha}\right\},
    \label{eq:mombalance}
\end{equation}
is derived along the centerline using assumed velocity fields which are scaled and modified by $s$ and $L$ in the projectile and $\alpha$ in the target of strength $Y_t$
We calculate $\alpha$ as a function of the interface velocity $u$, using the cylindrical cavity expansion,
\begin{equation}
    \left(1 + \frac{\rho_t u^2}{Y_t}\right)
    \sqrt{K_t - \rho_t u^2 \alpha^2}
    =
    \left(1 + \frac{\rho_t u^2 \alpha^2}{2 G_t}\right)
    \sqrt{K_t - \rho_t u^2},
    \label{eq:cylcavity}
\end{equation}
where $G_t$, and $K_t$ are the target's shear and bulk moduli.
The bulk modulus is further defined using a modified equation-of-state form,
\begin{equation}
	K_t = K_0 \left(1 + k \frac{u}{c_0}\right),
    \label{eq:eos}
\end{equation}
where $K_0$ is the elastic bulk modulus and $k$ is linear coefficient of the Mie-Gr\"{u}neisen shock velocity dependence on particle velocity.
Tail deceleration is derived from a differential approximation of elastic waves reflecting along the shaft of of the penetrator to give
\begin{equation} 
	\dot{v} = - \frac{\sigma_p}{\rho_p (L-s)}\left\{ 1 + \frac{v-u}{c} + \frac{\dot{s}}{c} \right\},
    \label{eq:tailrate}
\end{equation}
where $c$ is the longitudinal elastic wave speed of the projectile.
As there is a sustained difference in nose and tail velocities, we can easily define a rod erosion rate
\begin{equation}
    \dot{L} = u - v.
    \label{eq:lengthrate}
\end{equation}
The remaining equation to close the system is a continuous slope constraint on the velocity profile at the impact interface,
\begin{equation}
    s = \frac{R}{2}\left( \frac{v}{u} - 1 \right)\left( 1 - \frac{1}{\alpha^2} \right).
    \label{eq:smoothfield}
\end{equation}
This closes the solution in general, but supplying additional details can aid implementation.

Since these equations are to be time integrated, it requires posing this system as a coupling of first-order derivatives.
Equations \ref{eq:mombalance}, \ref{eq:tailrate}, and \ref{eq:lengthrate}, directly relate the rates, while \ref{eq:smoothfield} can be differentiated to do likewise.
By noting the cavity expansion equation \ref{eq:cylcavity} is quadratic in $\alpha^2$ when both sides are squared\cite{Walker2021}, we can solve for $\alpha$ in terms of $u$.
Differentiating the result with respect to time closes a linear system, offering well-posed for solution of the coupled rates for the system.
Numerical integration can then be performed using forward Euler with a sufficiently small timestep.

All that remains of the original model is the starting condition,
\begin{equation}
    P_{shock} 
    = \rho_p \left(c_p + k_p(u+v_0)\right) (u+v_0)
    = \rho_t (c_0 + k u) u
    \label{eq:shockjump1d},
\end{equation}
where $P_{shock}$ represents the shock pressure across the interface assuming a 1-D impact between the projectile and target materials.

\subsection{Parameter and Reference Model Elections}

From here we will soon discuss some modifications to this model, and for those demonstrations it will aid us to identify both a set of reference hydro-simulations and paired Walker-Anderson model parameters.
The hydro-simulation material properties and matched model parameters identified here will be used throughout the paper, unless we specifically state otherwise within that section.
For the purpose of providing detailed simulation reference data, we elect to make use of Lawrence Livermore National Laboratory's hydro-simulation code ALE3D \cite{noble2017ale3d}, run with fully Eulerian remap settings.
Within those simulations both our tungsten projectile and our steel target will be modeled by Johnson and Cook strength and failure models, with Mie-Gr\"{u}neisen equations-of-state; the specific material parameters are included in Appendix \ref{sec:params}.
Note: we make use of a set of models for tungsten and steel that suffice to include realistic values for most parameters; however, the final models are a composite of sources and have not been tuned to match specific test data.
  
\begin{figure}[H]
    \centering
    \fbox{\includegraphics[width=0.5\textwidth]{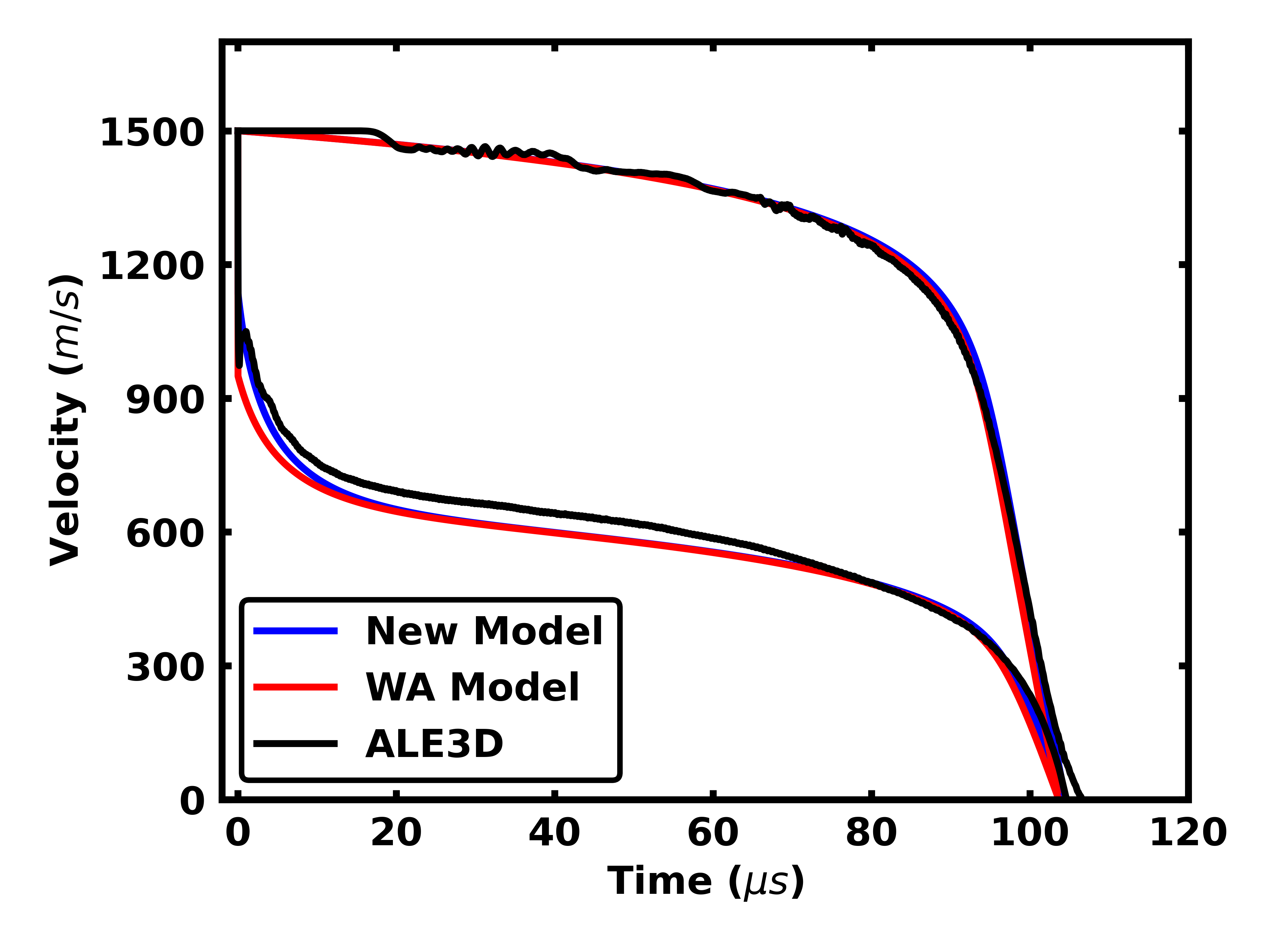}}
    \caption{Projectile nose and tail velocities of tuned Walker-Anderson model for 1500 m/s $v_0$ vs matched ALE3d results}
    \label{fig:basetune}
\end{figure}

For the purposes of our semi-analytic model, we retain the parameters of the original \cite{Walker1995} model and leave many unchanged in Table \ref{tab:matparam}.
\begin{table}[H]
\centering
\begin{threeparttable}[b]
\caption{Analytic model material constants \vspace{-5pt}}
\label{tab:matparam}
\begin{tabular}{rSSSSSSSS}  \hline \\[-8pt]
   &   $\mathrm{\boldsymbol{\rho}}$   &  $\mathrm{\mathbf{K}}$     & $\mathrm{\mathbf{E}}$     & $\mathrm{\mathbf{G}}$     & $\mathrm{\boldsymbol{c_0}}$ & $\mathrm{\mathbf{k}}$ & $\mathrm{\mathbf{Y}}$\\ 
   &   $\mathrm{\mathbf{(g/cm^3)}}$   &  $\mathrm{\mathbf{(GPa)}}$ & $\mathrm{\mathbf{(GPa)}}$ & $\mathrm{\mathbf{(GPa)}}$ & $\mathrm{\mathbf{(km/s)}}$ & $\mathrm{\mathbf{(-)}}$ & $\mathrm{\mathbf{(GPa)}}$\\ \hline \hline \\[-8pt]
\textbf{Steel}      & 7.85            & 166.7                  & 206.8                 & 76.9                  & 4.50                        & 1.49                  & 1.5{\textdaggerdbl}\\
\textbf{Tungsten}   & 17.00\tnote{\textdagger}  & 302.1        & 327.5                 & 124.1                 & 4.00\tnote{\textdagger}   & 1.24\tnote{\textdagger} & 1.5\tnote{*}   \\ \hline
\end{tabular}
\begin{tablenotes}
\item[*] Value from \cite{Chocron2003}
\item[\textdagger] Modified to match Appendix \ref{sec:params}
\item[\textdaggerdbl] Tuned via case in Figure~\ref{fig:basetune}
\end{tablenotes}
\end{threeparttable}
\end{table}
Except for the yield values $Y$, the modifications to the tungsten properties were made so that the analytic model was consistent with the ALE3D hydro-simulation model.
For the tungsten yield, we borrow a value determined previously \cite{Chocron2003}, leaving us only a single parameter to tune for the target strength.
To affect that tuning and provide a reference impact scenario, we employ the projectile velocity and geometry used in the development of the original half-space model \cite{Walker1995}.
This hemispherical headed projectile has a length-to-diameter ratio of 10, with a total length of 8.17 cm (0.817 cm diameter).
The results shown in Figure~\ref{fig:basetune} demonstrate our final target yield value of 1.5 GPa.
This family of parameters is used for all analytic models throughout this work regardless of the specific impact scenarios being considered, with specified modifications allowed for our experimental comparison.
The red curve demonstrates our ability to recover the performance of the original model for impact into an infinite half-space.
Within the next subsection we will modify the initial conditions for this impact, which are shown in blue.
As the updated model performs with only a slight difference over the length of the penetration no modification to the tuning is considered.

\subsection {Updated Initial Conditions}

\begin{figure}[H]
    \centering
    \fbox{\includegraphics[width=0.5\textwidth]{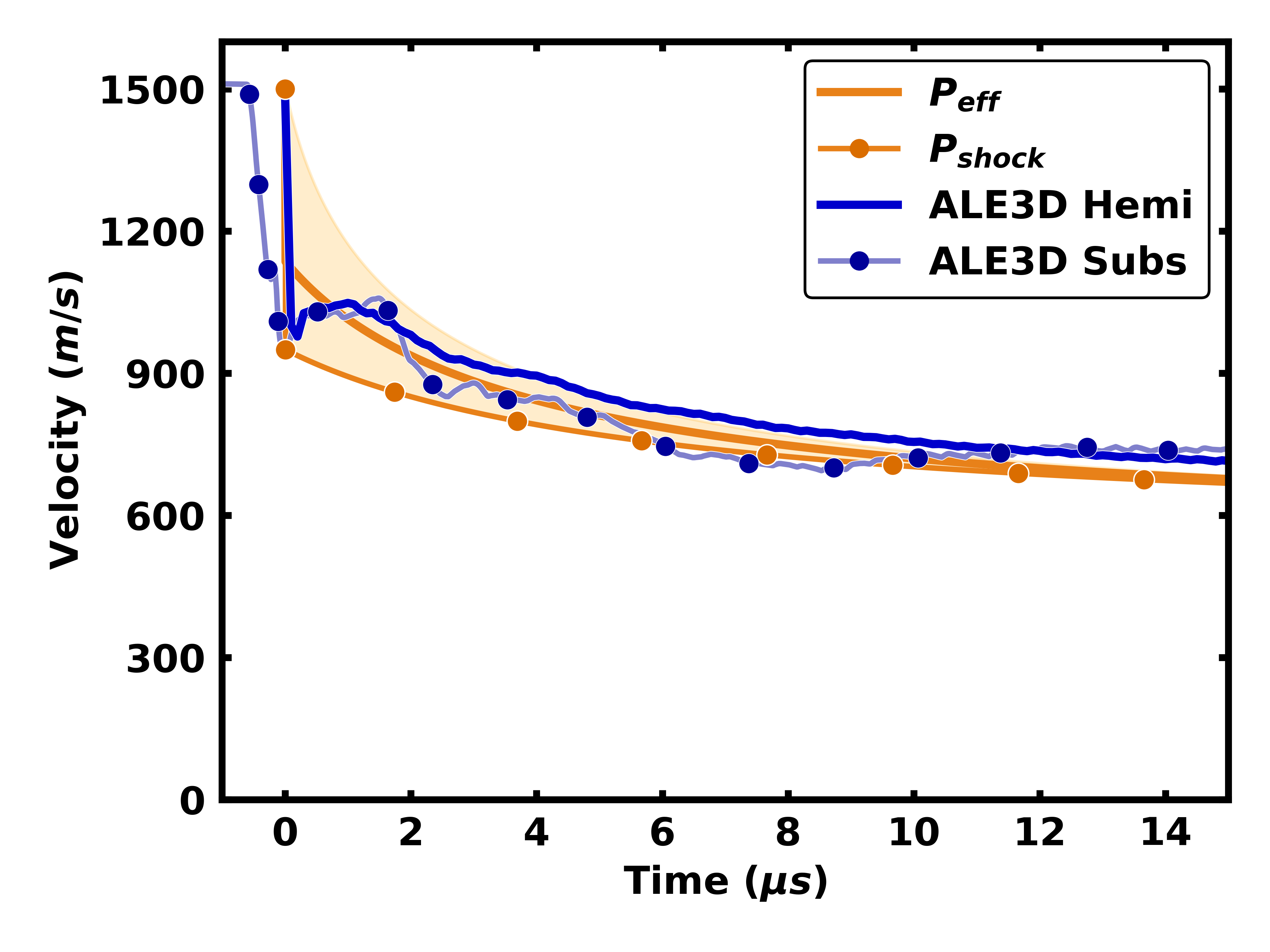}}
    \caption{Nose velocities of ALE3D hydro-simulations of hemi-headed and subsequent-impact scenarios(blue) and all possible analytic model initializations (orange). }
    \label{fig:initNose}
\end{figure}

Our next step is to update the starting conditions to better reflect a hemi-headed projectile over the 1-D shock equation found in (\ref{eq:shockjump1d}).
This condition is most correct for wide cylinder impacts, especially early in time at the center of the impact as one would expect for a 1-D shock calculation.
We did see in Figure~\ref{fig:thintests} that this does correctly model the initial drop in velocity at the projectile-target interface.
However, in both plotted cases, there is a quick rebound upward as the model transitions into steady state erosion.
To compare these effects for both initial and subsequent impacts, we introduce Figure~\ref{fig:initNose} to consider other initializations. 
The shaded region seen there represents the family of solutions found by assuming various shock initializations, from the top edge where the initial velocity is used directly to initialize the model, to the bottom edge that represents the traditional 1-D shock response.
Comparing against this range of possible considerations, we introduce two ALE3D simulation results in blue.
The thick blue line shows nose velocity history for a hemispherical head projectile, and the thinner line with circular markers shows response for a projectile that has already passed through a previous target before striking the considered half-space.
The subsequent impact case is time shifted to align with when the projectile nose crosses the plane of the target strike face, though advection welding in the ALE3D simulation causes the interface velocity to start dropping $\mathrm{\approx 0.75\ \mu s}$ early.
In general the subsequent impact does recover a slightly lower interface velocity than a comparable pristine flat cylinder, but the differences are not worth plotting.
Regardless, we see both ALE3D runs initially achieving the jump condition velocities shown in orange with circular makers, but with the rebound shown as discussed.
As a means to targeting this intermediate rebound we introduce the thick orange line that marks what we will call the \textit{effective-shock condition} which we will introduce shortly.

What we see in general is that the immediate impact aligns well with the 1-D shock assumptions, but that the irregularity of the flow following the impact causes the velocity to rebound, upward inside the bounded regions of initial assumptions.
We can define this family of initializations by introducing an additional constant, $C \in [0,1]$, into the initializing jump equation, 
\begin{equation}
    P_{C} 
    = \rho_p \left(c_p + k_p(v_0 - u)\right) (v_0 - u)
    = C\ \rho_t (c_0 + k u) u,
    \label{eq:arbshock}
\end{equation}
such that we can arbitrarily scale between our specified bounds.  We further introduce
\begin{equation}
    P_{eff}= \{P_C: C=\frac12\}.
    \label{eq:effshock}.
\end{equation}
as a tuned selection to match the plotted data.  

\begin{figure}[H]
    \centering
    \includegraphics[width=0.5\textwidth]{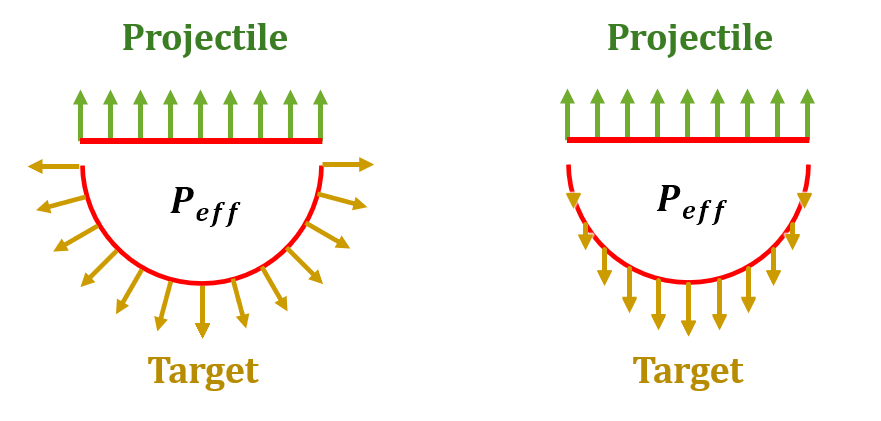}
    \caption{A visual sketch of the effective shock-jump condition (left) with the projection to axial contributions (right) }
    \label{fig:effshocksketch}
\end{figure}

\begin{figure}[H]
    \centering
    \includegraphics[width=0.65\textwidth]{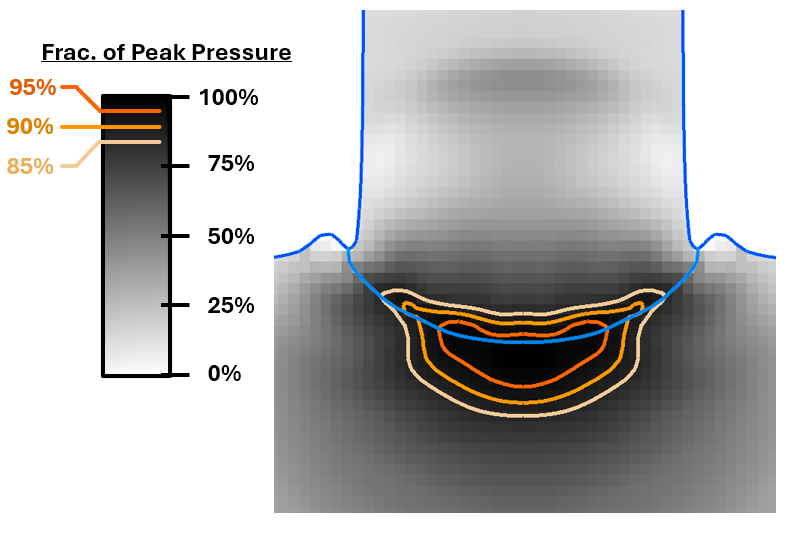}
    \caption{ALE3D pressure field of 1500 m/s hemispherical projectile and target 2 $\mu s$ after impact}
    \label{fig:ale3dpress}
\end{figure}

Seeking insights beyond an observational fit, we can build some intuition around this selected factor.
If we assume matched axial cross section, then this factor implies a non-matched curvature on the two sides of a finite shock region.
For simplicity if we further assume the projectile side of the shock is flat and the target side is hemispherical we arrive at a shock half bubble as sketched in Figure \ref{fig:effshocksketch}.
Our intuition here is that the flat side of the shock would correspond to the rod acting as a wave guide and the right represents the shock spreading in the target. 
On the left of the figure are the pressure fronts and their full normal vectors shown, with the axial projections of those normals sketched on the right.
Calculating the surface averaged axial component over the surface,
\begin{equation}
        C_{eff}=
        \frac{\int_0^{\pi/2}\cos(\theta)\sin(\theta)d\theta}
        {\int_0^{\pi/2}\sin(\theta)d\theta}
        =
        \frac12,
        \label{eq:avevec},
\end{equation}
with $\theta$ being the angle between the axis and the normal, we arrive at our observed factor.
Here we acknowledge two significant constructed assumptions we've compiled, both a jump condition targeting non-initial conditions, and a perfectly flat impactor pressure front coupled to a perfectly hemispherical pressure front.
As form of corroboration, if we take the hemisphere impact of our ALE3D sim reference in Figure \ref{fig:initNose} and plot the pressure field at 2 $\mu s$ we get the pressure field shown in Figure \ref{fig:ale3dpress}.
We can see there that the post impact pressure field does have features resembling our flat/hemisphere jump geometry, even if the field is not sharp enough to warrant a jump condition, and we can see how our modified jump condition would reasonable condition for this resulting field.
Ultimately this provides a reasonable grounding for our tuned factor of 1/2.

Looking back to Figure \ref{fig:initNose}, initialization using the $P_{eff}$ condition well-matches the pristine hemispherical head simulation and will be what we use for initial hemi-headed projectiles.
Likewise we similar rebound and decay in the non-pristine impact, but to nose velocities that are somewhat lower.
This is due, at least in part, to the higher strain rates of flat impacts strengthening steel before onset of thermal softening.
Since the model at present has no way to accommodate this hardening effect, for subsequent impacts in multi-target scenarios our model will use the original $P_{shock}$ condition to capture that additional erosion.

In a later section it will become clear for finite targets that a more detailed accounting of wave effects is required for our new starting condition.
One alluring idea is to make the model consistent with those next efforts by initializing the plastic domain extent by setting $\alpha=1$ and then expand $\alpha R$ at the bulk sound speed $c_0$. 
Walker does address this as a possibility in his book; however, in practice the $\dot{\alpha}$ drag term in (\ref{eq:mombalance}) then becomes significant enough that nose velocities drop even lower than the $P_{shock}$ condition while requiring much finer steps during time integration.
Our interpretation is that the drag implied by the $\dot{\alpha}$ term arises from an assumption that the flow field is smoothly varying in time, and this does not well-reflect early-stage crater development assumptions.
As such, we concede a discrepancy in the propagation of target flow that is instantaneous locally and will discuss its implications in the conclusion.

\subsection {Review of Finite Domain Model}

\begin{figure}[H]
    \centering
    \includegraphics[width=0.5\textwidth]{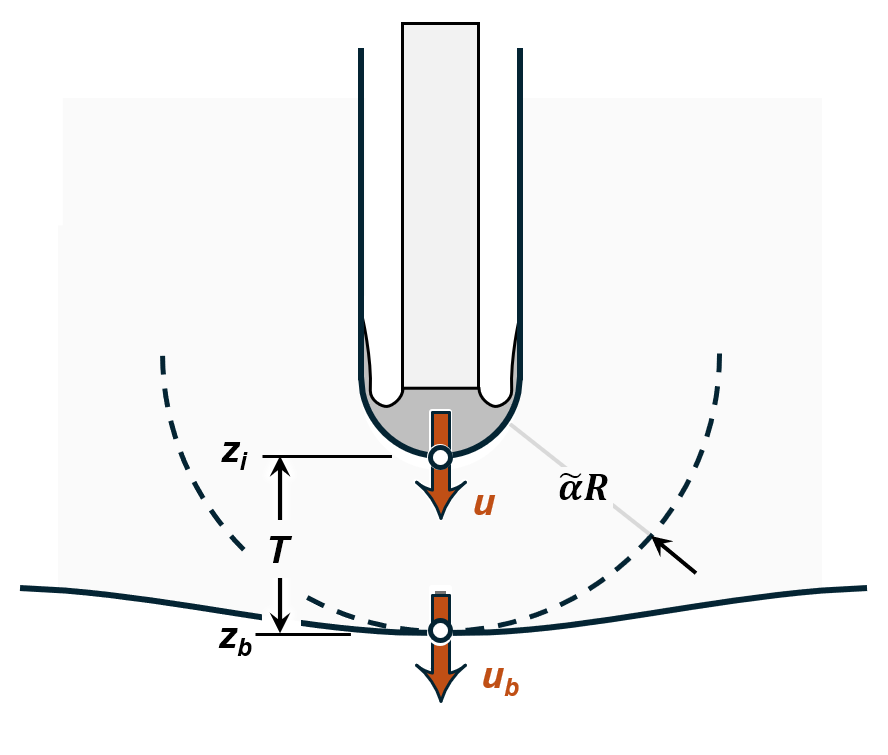}
    \caption{Added model parameters of projectile penetration into a finite target}
    \label{fig:backfacesketch}
\end{figure}

The Walker model for finite targets \cite{Walker1999bulge} modifies the assumed flow field and takes into account both the weakening target via shear area reduction, as well as material acceleration of the remaining unpenetrated plate material.
Again, we lay out the critical elements of that work to ease discussion of later-suggested improvements to the theory.
Relevant additional geometric details of finite back-face modeling are shown in Figure~\ref{fig:backfacesketch}.
There, we see the penetrator as it approaches the back-face with the sketch showing that the evolution of the back-face bulge.
When the plastic zone of the previous section reaches significantly beyond the back-face of the target,
\begin{equation}
    (T + R)^2 + R^2 < \alpha^2 R^2,
    \label{eq:transition}
\end{equation}
we transition to this finite model.
The variable $T$ has been introduced to articulate the distance between the projectile-target interface $z_i$ and back-face $z_b$.
This parameter will then be used to define a new parameter $\tilde{\alpha}$ that creates an $\alpha$-like parameter describing the proximity of the projectile of the back-face,
\begin{equation}
    \tilde{\alpha} = \frac{T + R}{R}.
    \label{eq:alphatilde}
\end{equation}
Likewise, we define a new radius parameter $\bar{R}$ describing the extent of the plastic zone given a still infinite target,
\begin{equation}
    \bar{R} = R(\alpha+1).
\end{equation}
These parameters then act as inputs to $\lambda$, a flow potential power weight between shallow and deep flow fields, calculated by
\begin{equation}
    \lambda = \frac{3T}{2\bar{R}}-\frac12\left( \frac{T}{\bar{R}} \right)^3.
    \label{eq:lambda}
\end{equation}
With that weighting calculated, we can then  calculate the centerline velocity of the back-face bulge.
\begin{equation}
    u_b = u \left( \frac{R}{T+R} \right)^{2\lambda}.
\end{equation}
Given these additional parameters of $\tilde{\alpha}$ and $u_b$, the final updates require only an update of drag terms in (\ref{eq:mombalance}), with
\begin{equation}
    \left\{\frac12\rho_t u^2 + \frac73 Y_t \ln{\alpha}\right\}
    \qquad \mathrm{being\ replaced\ by} \qquad
    \left\{\frac12\rho_t \left( u-u_b \right)^2 + \frac73 Y_t \ln{\tilde{\alpha}}\right\}.
    \label{eq:correctedmombal}
\end{equation}
It was with these relatively straightforward adjustments that the Walker back-face model was able to capture smooth re-acceleration of the projectile nose as seen in Figures \ref{fig:thicktest} where we see its improvement over the Ravid analytic model.
  
For clarity, we note that in the original model $\alpha$ and resultedly $\bar{R}$ were allowed to continue to evolve after encountering the back-face, however in our implementation both are frozen.
This was done because updating the radius for a theoretical plastic zone seemed unnecessary, though we do not observe any differences in the results.

\subsection{Limitations of the Starting Condition}

\begin{figure}[H]
    \centering
    \fbox{\includegraphics[width=0.5\textwidth]{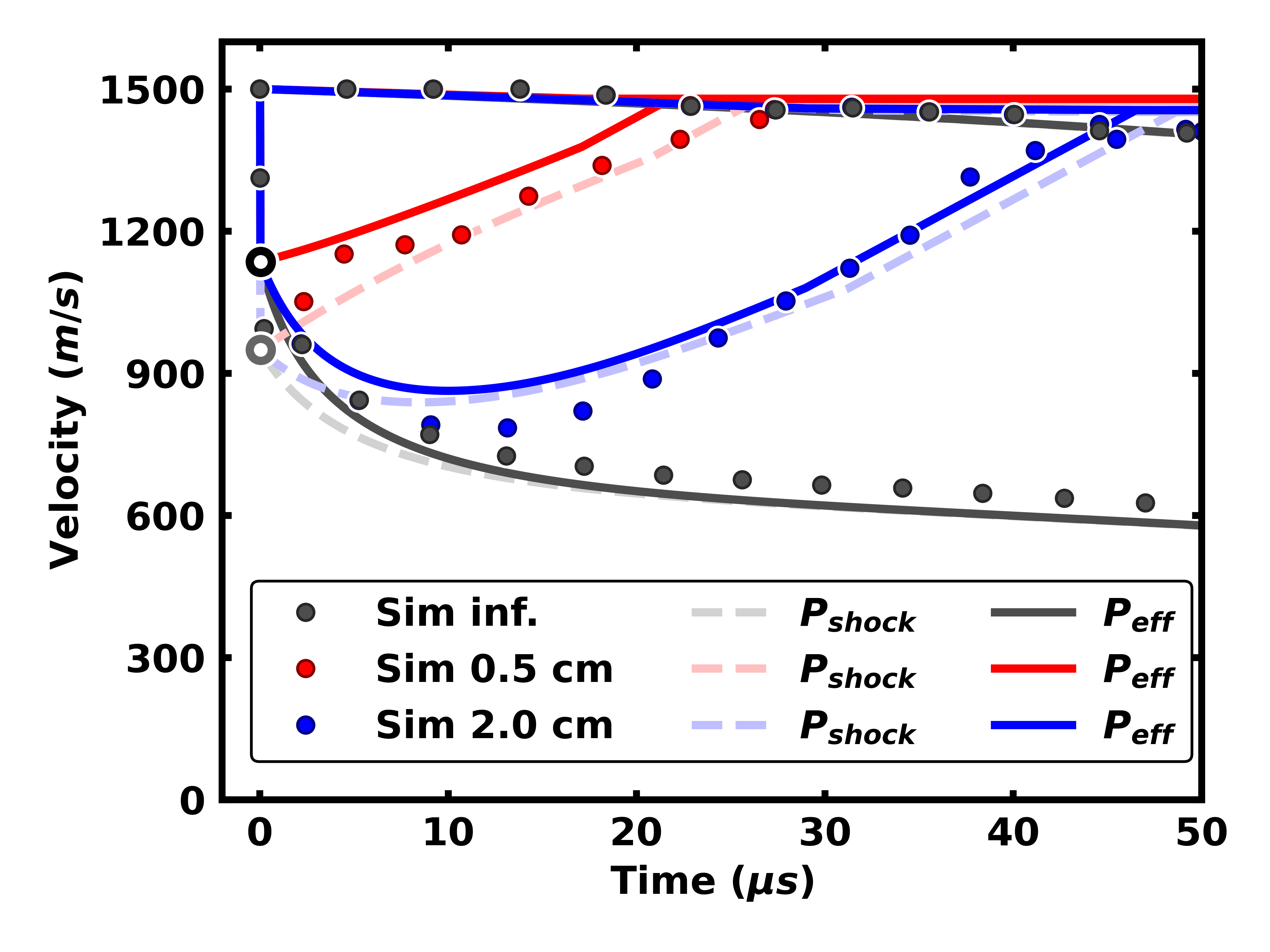}}
    \caption{Model velocity history of updated analytic model vs hydro-simulation }
    \label{fig:probeffshock}
\end{figure}

Having improved the starting condition for impacts into infinite targets, we naturally proceed to apply these improvements to finite targets with designs to improve back-face effects.
Unfortunately, our effective-shock initialization has a detrimental effect on our finite back-face predictions relative to the full shock used in the Walker model.
Figure~\ref{fig:probeffshock} shows results for  0.5 cm and 2.0 cm targets in red and blue with an infinite target results plotted in black.
Here, the dashed line for each case represents the Walker (shock) finite target response, while the solid line shows our new model with the effective-shock assumption (eff), and the color matched dots are samples from ALE3D solutions.
The gray and black dots with white centers mark the respective induced velocity of the 1-D and effective-shock assumptions.
Within the plot, we see that early time response for the infinite target shows the improvement discussed in the previous section as a local improvement over our reference solution.
As we examine the 2.0 cm target response it is apparent in Figure~\ref{fig:probeffshock} that effective-shock initialization does improve early time response and is able to match the Walker model in later time.
However, the 0.5 cm target underperforms relative to Walker, as the now-smaller initial drop in velocity is immediately followed by the target softening effects.

\begin{figure}[H]
    \centering
    \fbox{\includegraphics[width=0.5\textwidth]{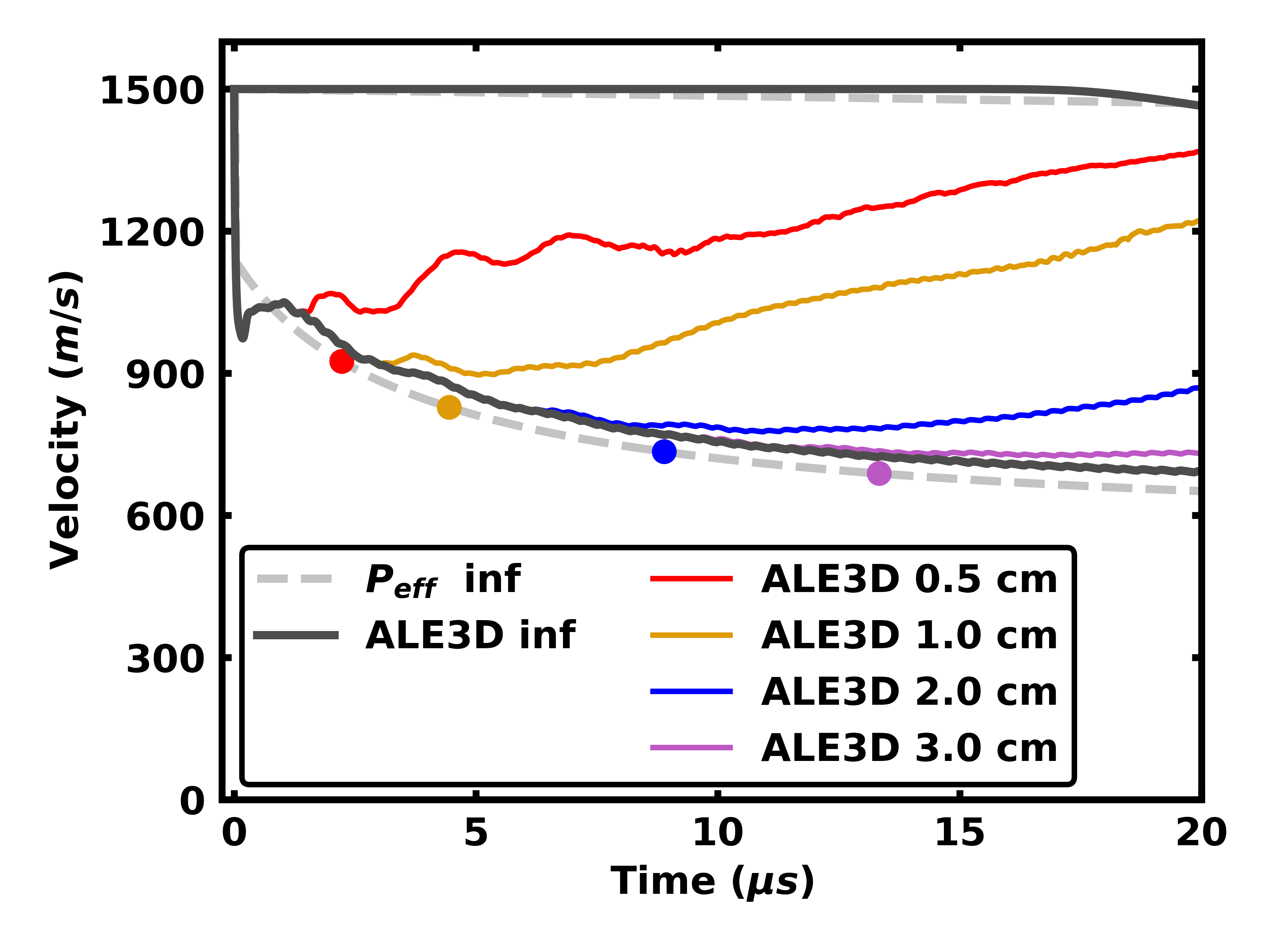}}
    \caption{Velocity history of updated infinite model compared against finite target ALE3D results }
    \label{fig:aleAgainstInf}
\end{figure}

This over-weakening of the thinnest target results from its back-face immediately being overrun by the plastic front that evolves around the crater.
The more dramatically the inequality of Equation \ref{eq:transition} is satisfied at initial time, the more we can expect an over-weakening of the target.
We do observe the same phenomena in the plots of Figure \ref{fig:thintests}, where the plots immediately start weakening for the Walker model, but we see here that 1-D shock assumption can balance this over-weakening to some extent.
In contrast, the plastic zone approaches the back-face more gradually in the 2.0 cm target and thus our modification is shown to be more reasonable.
To consider a possible correction, it is worth taking a moment to visually compare the ALE3D finite target curves against the infinite target solutions to see high similarities until the finite targets show target softening effects.
More specifically, we include Figure~\ref{fig:aleAgainstInf} which omits the finite target models completely and compares our early time effective-shock infinite model to the ALE3D results at various thicknesses.
Here, it becomes obvious that if we can properly delay the onset of finite target effects, we will be guaranteed better agreement up to that departure point.
In some sense, Ravid's style of infinite-until-failure \cite{Ravid1998,Chocron2003} modeling attempts to capture this effect through strain failure estimates, but Walker's later work elegantly showed that velocity recovery of the nose does not directly imply target failure.
Suspecting that wave transit time is required, we can naively estimate forward-and-back transit time using the target's bulk sound speed $c_0$ and the full target thickness.
Since the projectile's progressing penetration would reduce return times this serves only as an upper bound estimate, with larger error for thin targets.
Now plotting our infinite model interface velocity at those times, we get the color-matched dots shown in Figure~\ref{fig:aleAgainstInf}.
Clearly these dots do not perfectly capture the divergence of finite and infinite responses, but given the coarseness of the estimate we intuit a need for further investigation of target wave transients.

\subsection{Intuition from Detailed Simulations}

To better characterize wave transients within the target, we will observe how such effects evolve within more detailed ALE3D simulations.
These simulations employ a shock capturing method that resolves non-linear wave transients without the need to explicitly track their location.
For the purpose of identifying wave characteristics, we have placed 40 Lagrangian tracers on the centerline of a 2D simulation of a 1 cm thick target.
Each tracer outputs position, velocity, and effective plastic strain histories.
From these histories we can readily generate forward traveling characteristics by looking at initial rises in velocity and plasticity.
However, since we are specifically trying to capture finite target effects, it becomes necessary to create a second infinite target simulation with matched initial tracer locations.
With that second simulation, we can then observe how the finite target velocities diverge from the half-space velocities to construct a reflection characteristics from the back-face. 
  
\begin{figure}[H]
    \centering
    \begin{subfigure}[t]{0.48\textwidth}
        \centering
        \fbox{\includegraphics[width=0.98\textwidth]{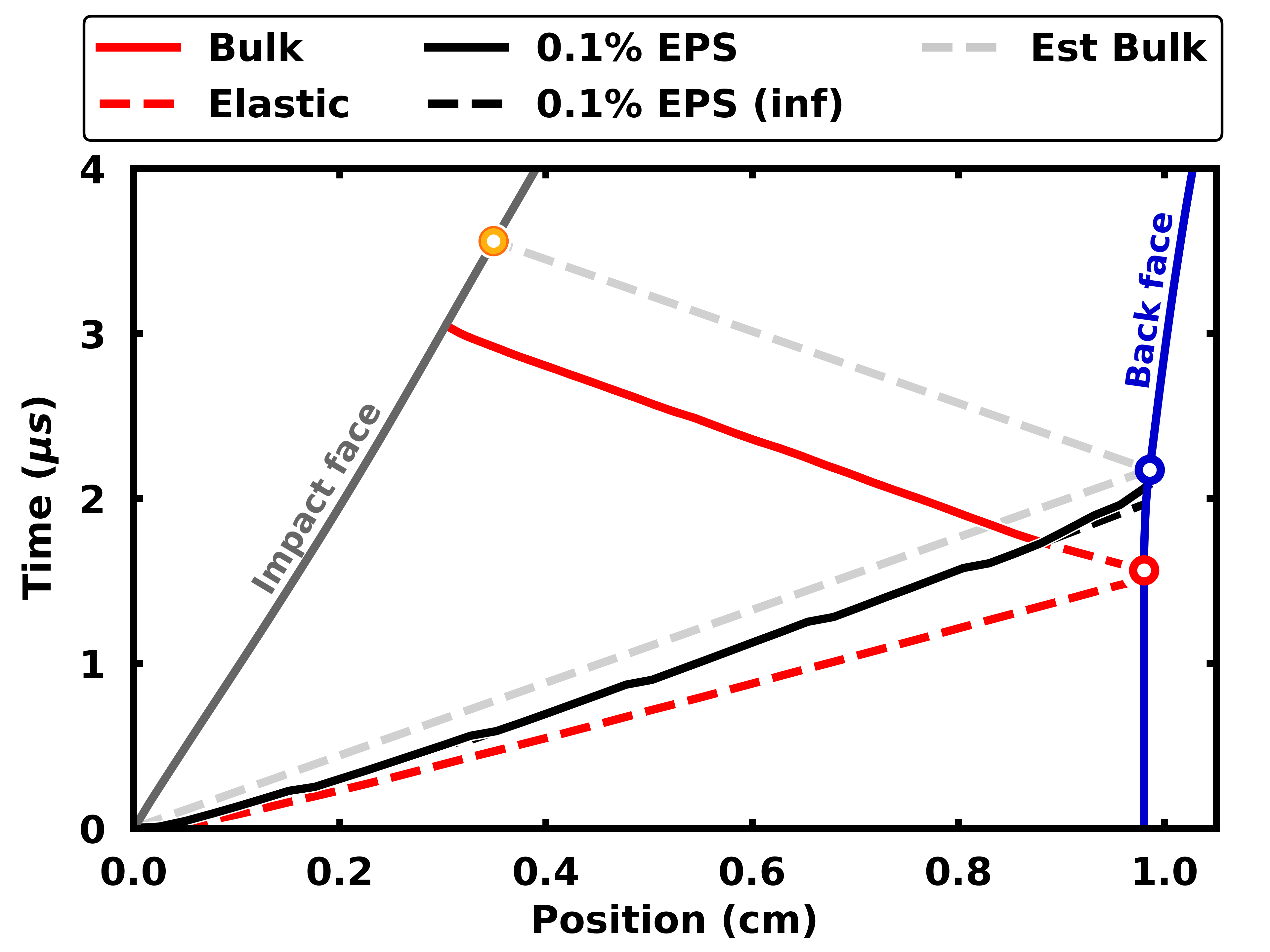}}
    \caption{x-t diagram propagated effects}
        \label{fig:aleXTdiagram}
    \end{subfigure}
    ~
    \begin{subfigure}[t]{0.48\textwidth}
        \centering
    \fbox{\includegraphics[width=0.98\textwidth]{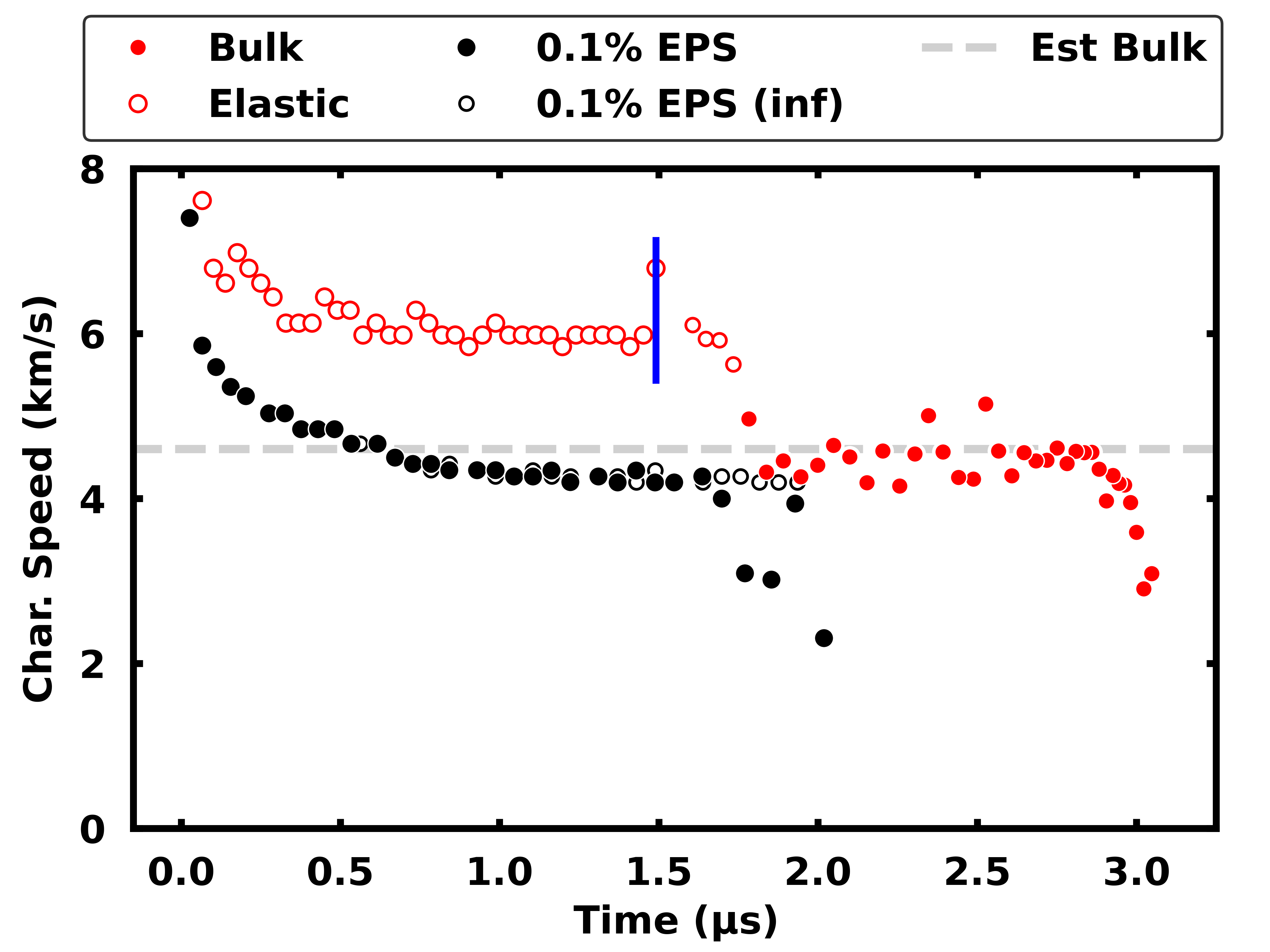}}
    \caption{Speed of wave propagation in time}
        \label{fig:aleXTslopes}
    \end{subfigure}
    \caption{Summaries of 1cm velocity and plasticity characteristics as extracted from ALE3D simulations. Reds indicates velocity characteristics, blacks indicates plasticity characteristics and dashed grey indicates a bulk speed assumption. Dark grey marks the front face, and blue indicates the back-face, or the arrival characteristic at the back-face.}
    \label{fig:alecharplots}
\end{figure}

The results of this process are shown in Figure~\ref{fig:alecharplots}, with Figure~\ref{fig:aleXTdiagram} being the x-t diagram plotting the characteristic's path-time histories and Figure~\ref{fig:aleXTslopes} providing matched numerical derivatives of those curves.
Given that curves were derived from rise times and that the tracers are advected through the domain, the resulting numerical derivatives are somewhat noisy.
Here, the red curves and markers denote the characteristics generated from velocity rises and differences, with red dashed lines and hollow markers indicating response when the domain is elastic.
Since these curves and markers trace the propagation of impact effects both to and from the back-face, we will refer to the joint entity as the impact characteristic.
In Figure~\ref{fig:aleXTslopes}, a vertical blue line is used to mark when the red hollow dots transition between forward and backward characteristics.
The black lines and markers denote the progression  of the plasticity front through the domain, with dashed lines and hollow dots indicating the infinite target response most visible after the plastic front crosses the backward characteristic.
Finally, because we are most concerned with plastic wave propagation, we append the dashed light grey lines within both plots to represent a bulk sound wave propagated, forward and then backward, from the impact time.

Starting at the origin and proceeding in time, we see over driven shock speeds in both the red impact and black plastic characteristics caused by the projectile induced material velocity.
Both fronts then decay with propagation to the longitudinal and bulk wave speeds.
As the impact characteristic reflects off the back-face and returns, it continues to move at the longitudinal wave speed, until it transitions to the bulk wave speed as it enters the plastic zone.
Once there, we see the wave remains steady until the now on-coming material velocity drives down the shock speed in the laboratory frame.
Ultimately, Figure~\ref{fig:aleXTslopes} shows that the bulk sound speed represents the best mean approximation of the propagation of plastic effects.
We find this to be true in the direct sense that it matches the steady propagation speeds, and in the more general sense that even forward and backward shock effects will tend to cancel about this value.

\begin{figure}[H]
    \centering
    \fbox{\includegraphics[width=0.5\textwidth]{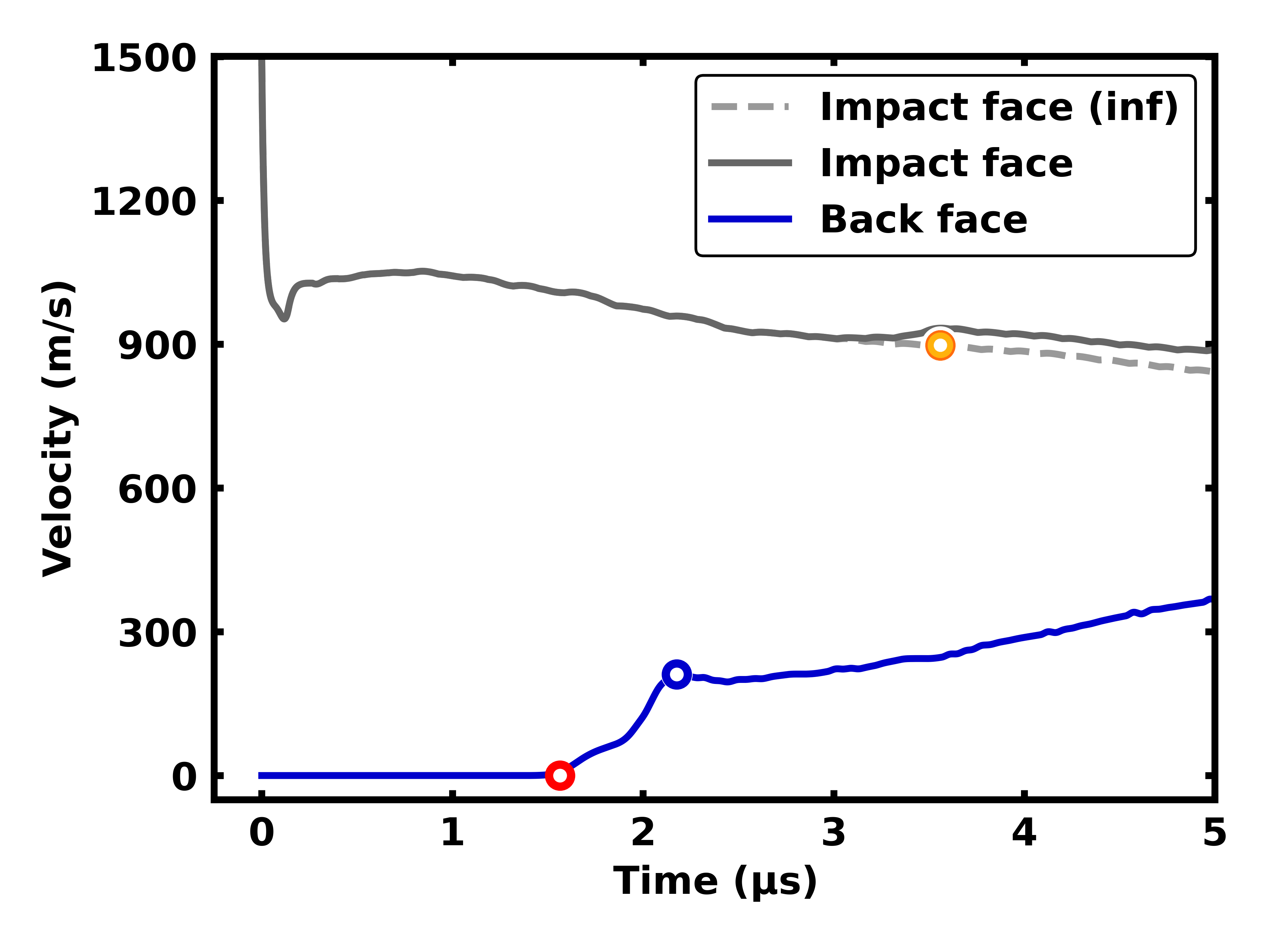}}
    \caption{Velocity history of ALE3D simulation results denoting arrival times of observed characteristics and estimates.  Red is the elastic arrival at the back-face, blue and orange are the respective estimates of arrival using the bulk wave speed.}
    \label{fig:aleVTdiagram}
\end{figure}

As a final check on our ability to use the bulk sound speed, we plot velocity histories of the target faces as seen in Figure~\ref{fig:aleVTdiagram}.
There, we mark the points of interest within the velocity histories with hollow, colored dots, with each discussed point having a matched dot within the figure and in the earlier Figure~\ref{fig:aleXTdiagram}.
On the blue curve plotting the back-face history, the red dot marks the velocity when the impact characteristic arrives at the back-face, while the blue dot marks the velocity when the initial impact is propagated to the back-face by the bulk sound speed.
These two marks show that the simplified propagations reasonably estimates the time we would associate with the start of steady plastic motion of surface.
On the grey dashed curve, plotting interface history for an infinite target, the orange dot marks when the impact reflection is estimated to arrive using the bulk sound speed.
We do see that the solid grey line of the finite target interface diverts from the infinite earlier than our marker indicates; however, our calculated point is most adjacent to a visible kink in the finite target history, which marks when the two curves begin moving apart in earnest.
This highlights how propagating target effects using $c_0$ reasonably estimates important features in the velocities of the target faces, and we now expect that accounting for updated projectile location will correct the late arrival times estimated in Figure~\ref{fig:aleAgainstInf}.

\subsection{Wave Corrections}

With our simulation based intuitions of propagating effects firmly in hand, we now go about modifying the model. 
We start by borrowing from characteristic forms of the 1D wave equation by writing,
\begin{equation}
    \begin{split}
      K^+ &= z + \ c_0 t = z^+ + \ c_0t^+ \qquad \mathrm{and} \\
      K^- &= z - \ c_0 t = z^- - \ c_0t^- ,
  \end{split}
    \label{eq:chardef}
\end{equation}
where constant values of $K^+$ and $K^-$ define forward and backward traveling characteristics in $z$ and $t$ along which information propagates.
However, rather than operate on the $K$ values, we equivalently make use of characteristic times $t^+$ and $t^-$, which correspond to when a given characteristic crosses reference coordinates $z^+$  and $z^-$.
Respectively, we select both of these coordinates to be the initial front face location of the target, though the specific locations do not matter.
With those characteristic times we can now link properties of the front and back faces while affecting wave delays.

Given our excellent agreement for the infinite half-space target, we will restrict our modifications to the finite target terms, implying that all unmodified terms associated with the target live at the projectile interface.
As we think about the drag terms as they are applied at the interface,
\begin{equation}
    \frac12\rho_t \left( u-u_b \right)^2 + \frac73 Y_t \ln{\tilde{\alpha}},
    \tag{\ref{eq:correctedmombal}}
\end{equation}
our task now requires us to identify what properties need to be propagated from which locations across the target.
Clearly, back-face velocity $u_b$ would need to be carried along the negative characteristic from the back-face, while $\tilde{\alpha}=\tilde{\alpha}(z_i,z_b)$ and would require the interface position and the propagated back-face location.
Given our propagation via the negative characteristic of time $t^-$ then we can write
\begin{subequations}
    \begin{align}
        u_b &= u_b(t^-) \qquad \mathrm{and} \\
        \tilde{\alpha} &= \tilde{\alpha}(z_i,z_b(t^-)).
    \end{align}
\label{eq:negchar}
\end{subequations}
We now relocate via our $t^-$ characteristic time to the back-face and the calculation of both $z_b$ and $u_b$, where the negative characteristic intersects a forward characteristic defined by $t^+$.
Back-face location $z_b$ requires only integration of the velocity
\begin{equation}
    z_b = \int_0^{t_b} u_b dt,
    \label{eq:bfacecalc}
\end{equation}
where $t_b$ is the time when the characteristic crosses the back-face.
Recalling the form of the back-face velocity itself, $u_b = u_b(z_i, u, z_b)$, together we again borrow propagated properties from earlier in time, now using the positive characteristic giving
\begin{equation}
	u_b = u_b\left( z_i(t^+), u(t^+), z_b, \alpha(t^+) \right).
    \label{eq:poschar}
\end{equation}
Aside from the delays implied by this form, we retain the decay of velocity in space as originally derived by Walker \cite{Walker1999bulge}.
This coveniently preserves mass conservation of the assumed flow field, as long as its considered within the context of propagated effects.
As a minor note, the expanded form of $u_b(z_i, u, z_b)$ requires the calculation of $\lambda$ from $T$ as expressed in (\ref{eq:lambda}), which for clarity we highlight as distinct from the $T$ used to calculate $\tilde{\alpha}$ across the negative characteristic.
  
\begin{figure}[H]
    \centering
    \includegraphics[width=0.5\textwidth]{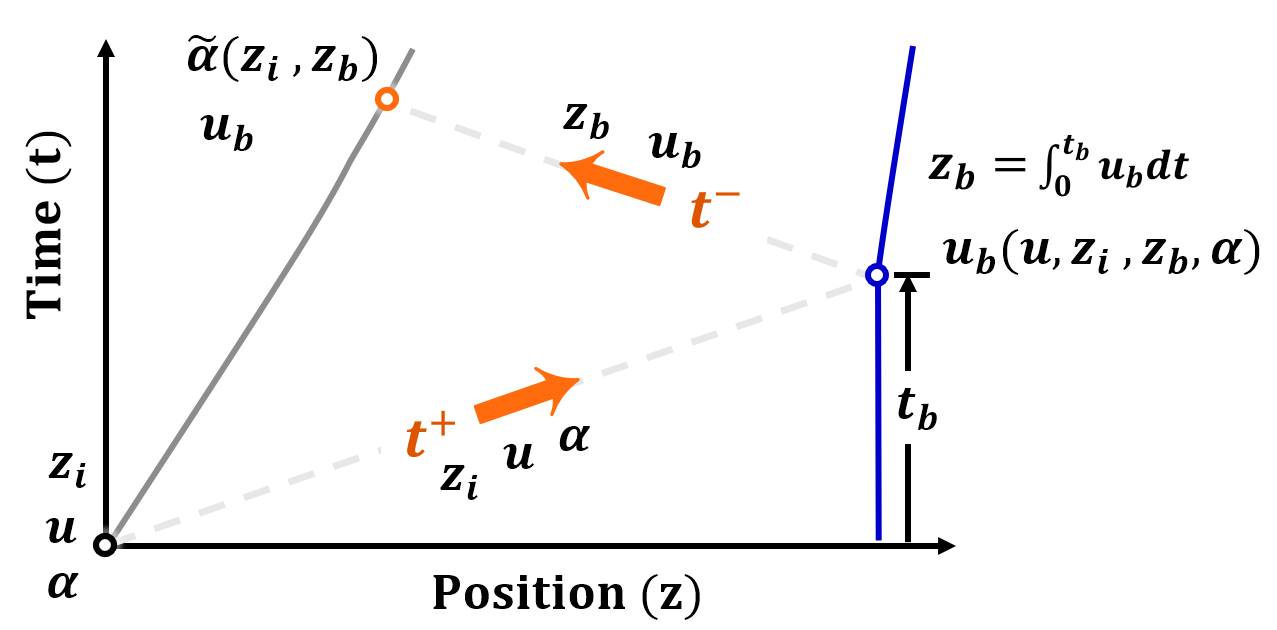}
    \caption{Property propagation via characteristics }
    \label{fig:wavepropsketch}
\end{figure}
We include the diagram of Figure~\ref{fig:wavepropsketch}, to make apparent how the relevant properties are propagated forward in time to future states that live on matched characteristics.
As we examine the flow of information, it becomes apparent that while the $u_b$ used at the final time requires both forward (via $u$) and backward propagation, the $\tilde{\alpha}$ would begin influencing the interface only after a backward propagation.
To correct this effect, we append the constraint that the model switch from using the $\alpha$ of the infinite half-space to the finite target $\tilde{\alpha}$ only after the arguments of the drag term receive a non-zero back-face velocity.
The result is as intended, where, until the time has passed for both directions of transit, the model is identical to infinite model.

For the reader's convenince, our previous sections have supplied the equations and implementation details necessary to fully reproduce the models \cite{Walker1995,Walker1999bulge} upon which our extension is directly based. 
While such an implementation-from-scratch would not be without effort, the extension to add wave delays would be comparably light work.
To demonstrate, we append Algorithm \ref{alg:phase_delay_short} to the expressions and articulations of this section.

\begin{algorithm}[H]
\caption{Phase-Delayed Front-Back Face Coupling}
\label{alg:phase_delay_short}
\begingroup
  \renewcommand{\baselinestretch}{1.2}
  \selectfont
  \begin{algorithmic}
  \STATE \textbf{Inputs:} $t$, $z_i$, $u$, $\alpha$
  \STATE Compute backface characteristic time $t^+_{\mathrm{query}} \gets t - \frac{z_b - z_0}{c_0}$
  \STATE Interpolate $u^+$, $z_i^+$, $\alpha^+$ at $t^+_{\mathrm{query}}$ from history
  \IF{plastic zone using $u^+$,$\alpha^+$,$z_b$ has reached back face}
  \STATE Compute current back-face velocity $u_b$ using $u^+$, $z_i^+$, $\alpha^+$
  \STATE Update current back-face position: $z_b \gets z_b + u_b \Delta t$
  \ENDIF
  \STATE Store current nose/interface state $(t^+_{\mathrm{new}}, u, z_i, \alpha)$ in forward history
  \STATE Store current back-face state $(t^-_{\mathrm{new}}, u_b, z_b)$ in backward history
  \STATE Compute front-face characteristic time $t^-_{\mathrm{query}} \gets t + \frac{z_i - z_0}{c_0}$
  \STATE Interpolate $u_b^-$, $z_b^-$ at $t^-_{\mathrm{query}}$ from backward history
  \IF{$u_b^- > 0$}
  \STATE Use $u_b^-$, $z_b^-$ to update drag terms
  \ELSE
  \STATE Use infinite-target values
  \ENDIF
  \end{algorithmic}
\endgroup
\end{algorithm}

As we can see, the core concept is not complicated, with information recorded at characteristic times and later referenced via interpolation of adjacent characteristics.
Aside from some cost for storing a detailed history, the method is not expensive as the interpolation can advance from the previous step's interpolation index to quickly update relevant history without significant search costs.
Furthermore, this strategy retains previous-model performance for thick targets because the later-time velocity changes during deeper penetrations is steadier, thus reducing the effect of a phase delay.

\begin{figure}[H]
    \centering
    \fbox{\includegraphics[width=0.5\textwidth]{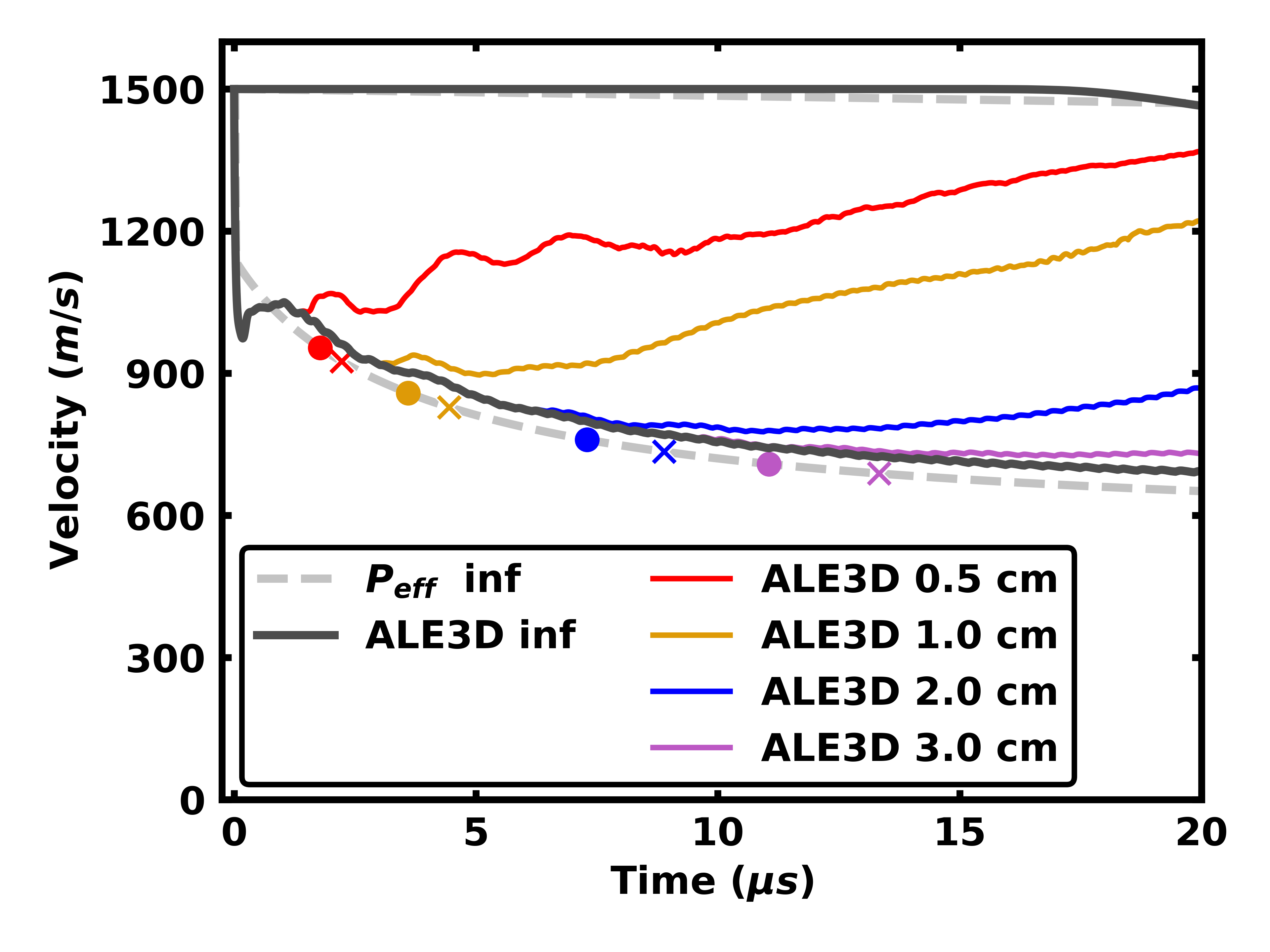}}
    \caption{The updated infinite model compared against finite target ALE3D results showing naive (x marks) and wave modeled (solid circles) estimates of transit times}
    \label{fig:aleAgainstInf2}
\end{figure}

Our first verification of the implementation, shown in Figure \ref{fig:aleAgainstInf2}, is a check the modeled arrivial times against the naive estimates used to generate Figure \ref{fig:aleAgainstInf}.
Here our previous estimates are shown with x markers, while the updated estimates are shown with solid circles.
As we would expect, our transit times are noticeably improved to better reflect how penetration reduces the information propagation across the target.

\begin{figure}[H]
    \centering
    \fbox{\includegraphics[width=0.5\textwidth]{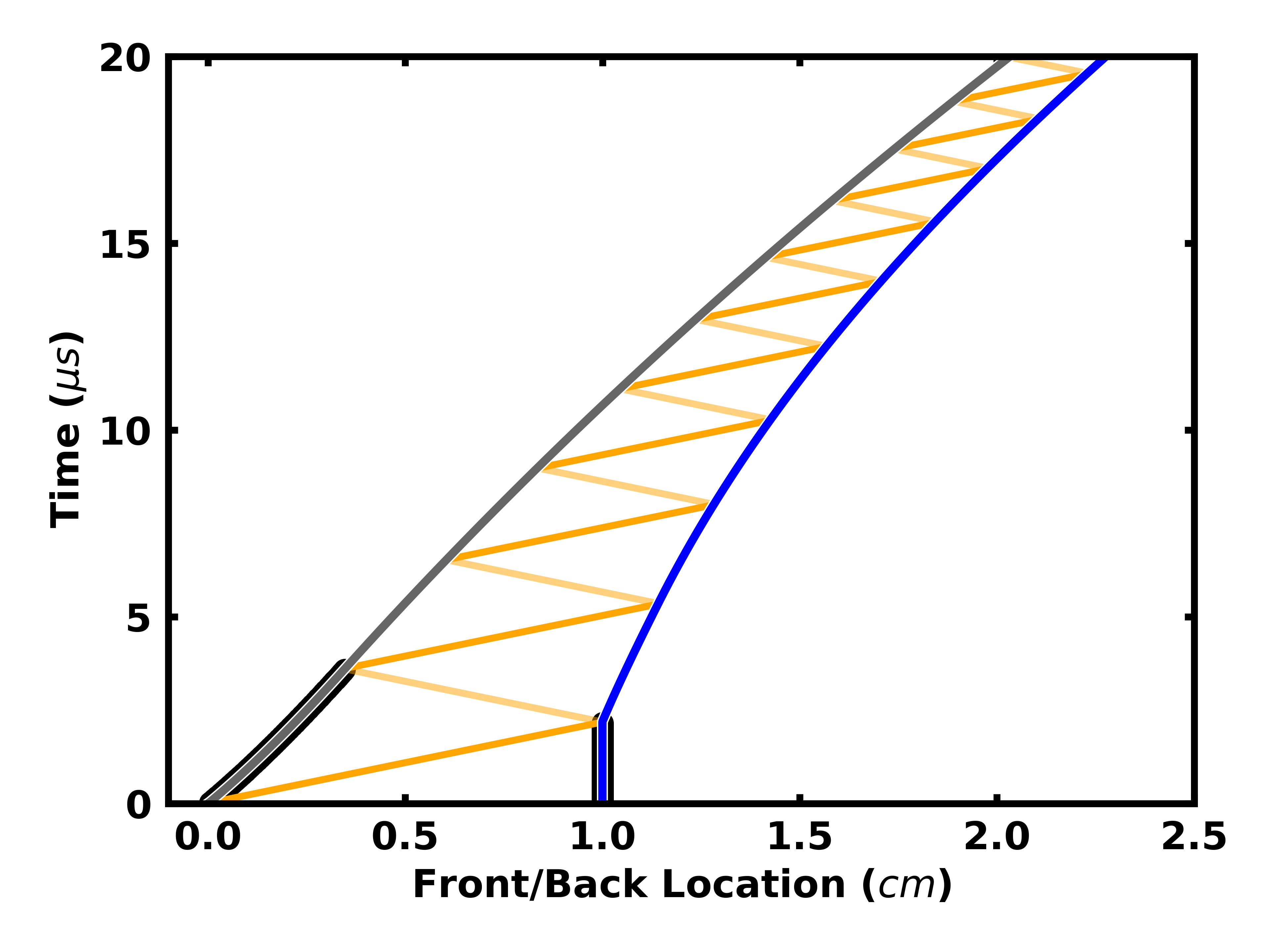}}
    \caption{An x-t diagram of a 1 cm target impact, showing forward and backward wave reflections dark and light orange.  Dark grey and blue mark the front and back target faces, and the black lines indicate face movement unaffected by the repective opposite faces. }
    \label{fig:xtwavereflect}
\end{figure}
Additionally, to verify proper behavior across the finite/infinite model transition Figure~\ref{fig:xtwavereflect} shows an x-t diagram where the progression of reflections from the initial impact are plotted.
As earlier, the projectile interface is shown in grey and the back-face is shown in blue, with orange lines tracing the impact characteristic as it moves forward (dark) and backward (light) across the target.
Thick black lines additionally highlight the part of the surface histories where the associated surface is unaffected by the other face of the projectile.
Aside from verifying implementation and the successful delay of finite target effects, we also see that as the target thins, the reduction in phase delay is captured.

\subsection{Target Failure Modeling} \label{sec:nofail}

Previous back-face models have specifically addressed the need to model projectile velocities after target failure \cite{Chocron2003,Walker2021} as a distinct phase.
Both models shown earlier in Figure~\ref{fig:thicktest} include such effects, with them being essential in the Ravid model.
However, for both our updated model, and Walker's back-face model, specifically triggering a projectile recovery reduces our ability to match our ALE3D simulations.
This occurs because the effects of a target's late-term strenth are very small, and the natural nose re-acceleration captured by the mechanics is a more rigorous modeling of recovery than those used after failure.
Within our work, the projectile is considered recovered once the nose velocity matches the tail velocity, or if a subsequent impact needs to be initiated.
This is caused, at least in part, by fact that target drag due to strength is low as the projectile passes towards the back-face and projectile momentum's domination of the late result means the finite target balance naturally models recovery.
We note that using the multi-material zones within hydrodynamic simulations can condition our baseline simulations to over-state ductile response; however, Section \ref{sec:expvalid} will provide an experimental comparison that validates our assumption.
If considering such modeling, we suggest the updated recovery form \cite{Walker2021} over the original \cite{Chocron2003}.

\section{Experimental Validation}\label{sec:expvalid}

Before proceeding to thin-target predictions, which make significant use of detailed simulations as their reference, we first present an example demonstrating model efficacy against experimental results.
For this purpose, we turn to available data for finite targets \cite{AndersonStilp1995}.
This data comes in the form of position-time data for an impact near $V_{50}$ that we can use for calibration, as well as exit velocities across impact speeds for validation.
All experimental setups used for this comparison feature blunt-nosed tungsten rods with an aspect ratio of 12.5 and a length of 5 cm impacting a 2.90 cm-thick steel target.
Validation against this data set is sufficient to demonstrate the continued efficacy of analytic models against finite targets and is of particular interest given our election not to include an explicit recovery phase in our code.

\subsection{Material Calibration}

The experiments supplying position-time data for this comparison centered their impact speeds on the $V_{50}$ of 1250 m/s to maximize interactions with the back-face of the target.
Their x-ray images across similar tests had different timing delays allowing for them to construct composite time-position data for the projectile nose and tail.
The authors also provide a post-test image of a non-peforated specimen which impacted at 1240 m/s (test number 4812), from this image the final depth of nose penetration is estimated to be between 2.55 to 2.60 cm.

\begin{figure}[H]
    \centering
    \fbox{\includegraphics[width=0.5\textwidth]{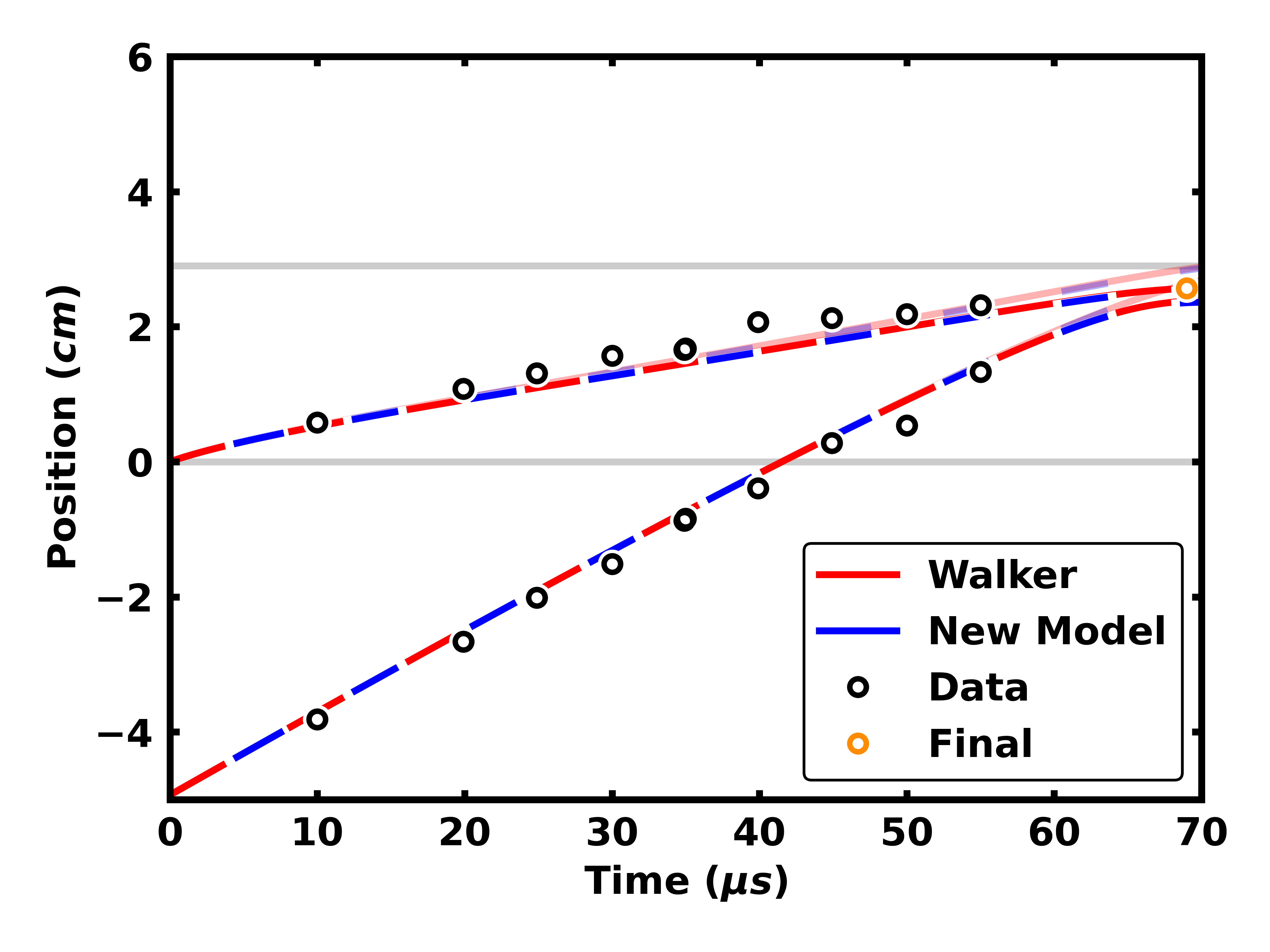}}
    \caption{An x-t history comparison for 2.90 cm target data taken in a series of tests at, or near, 1240 m/s.}
    \label{fig:xtdatacalib}
\end{figure}

To model this system using the analytic models, we make use of the same material parameters as discussed previously, with two exceptions.
The first is to use the original shock jump condition to better model the blunt projectile nose, and the second is to use alternate strengths for projectile and target.
We note that static yield values are supplied for both projectile and target, with the respective values being 1.2 GPa and 1.45 GPa, but dynamic yield values for our penetration models can differ significantly.
Since we lack an independent way to tune the target and projectile yields, our analytic models make use of the tungsten model yield of 1.2 GPa and modify the target yield such that it matches the available data.
We note that agreement with the position-time points was reasonably good with the supplied target yield value; however, the final penetration was almost 2.9 cm, which is at odds with the observed results.
However, by raising the target yield to 1.54 GPa, we are able to reproduce both the data points and obtain a reasonable final depth of penetration, as shown in Figure~\ref{fig:xtdatacalib}.
Here, we see the data points from our source, with an additional point in orange marking the final penetration depth, compared against the two analytic models.
Our plot also includes, in matched lighter colors, the results of the original strength parameters for reference.
We note here that both analytic models perform similarly because the targets tested here are thick and wave delay effects within the target remain small.
In general, we see that this limited tuning is well-able to match the available data, and we can now assess model performance against exit-velocity measurements.

\subsection{Results Validation}

\begin{figure}[H]
    \centering
    \fbox{\includegraphics[width=0.5\textwidth]{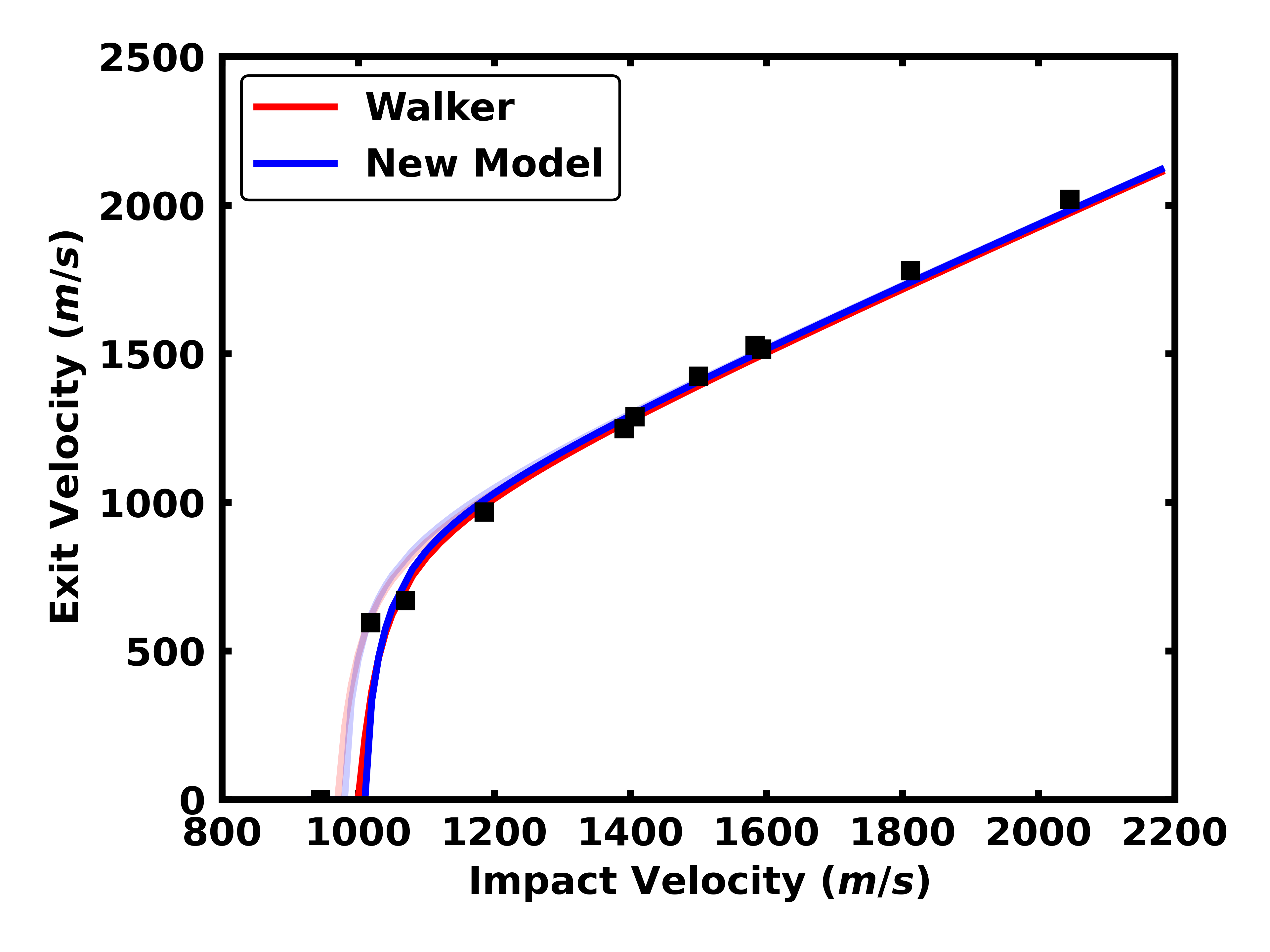}}
    \caption{Exit velocities for a 2.90 cm target over a range if impact velocities.}
    \label{fig:experimentvalid}
\end{figure}

Moving from calibration to validation, Figure \ref{fig:experimentvalid} compares the experimental exit velocity data against the results of our analytic models.
In general, this level of agreement is impressive, as the curves directly capture both the $V_{50}$ transition and the velocities after recovery.
As before we've also included unmodified strength values in light matched colors, and we see the results in this plot are not dramatically different showing the results are not overly sensitive to tuning.
More explicitly it captures both these outcomes without needing to specifically account for target failure.
In Subsection \ref{sec:nofail} we noted that relative to detailed simulations, explicitly omitting the failure criteria within the analytic models provides better agreement.
Figure \ref{fig:experimentvalid} here raises that model choice from a possible artifact of numerical advection to an experimentally observable outcome for both analytic models within this impact regime.
Unfortunately, as during calibration, this particular experimental data set does not let us observe the analytic performance of our models against thin targets.
Lacking similar data for thin plates we turn to detailed ALE3D simulations for such comparisons.

\section{Results} \label{sec:Results}

\begin{figure}[H]
    \centering
    \begin{subfigure}[t]{0.48\textwidth}
        \centering
        \fbox{\includegraphics[width=0.98\textwidth]{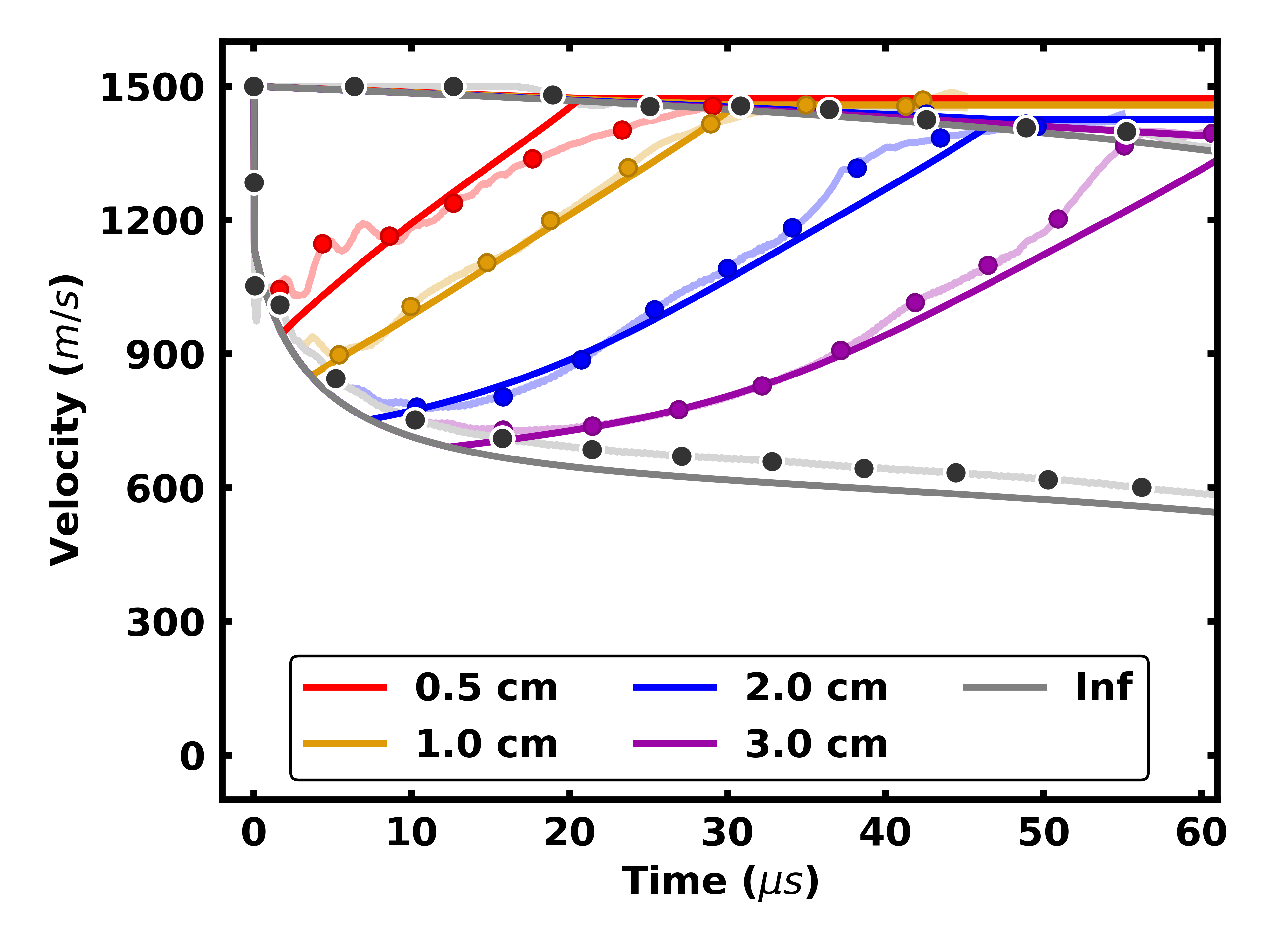}}
        \caption{Updated Model vs ALE3D}
        \label{fig:walkerWave}
    \end{subfigure}
    ~
    \begin{subfigure}[t]{0.48\textwidth}
        \centering
        \fbox{\includegraphics[width=0.98\textwidth]{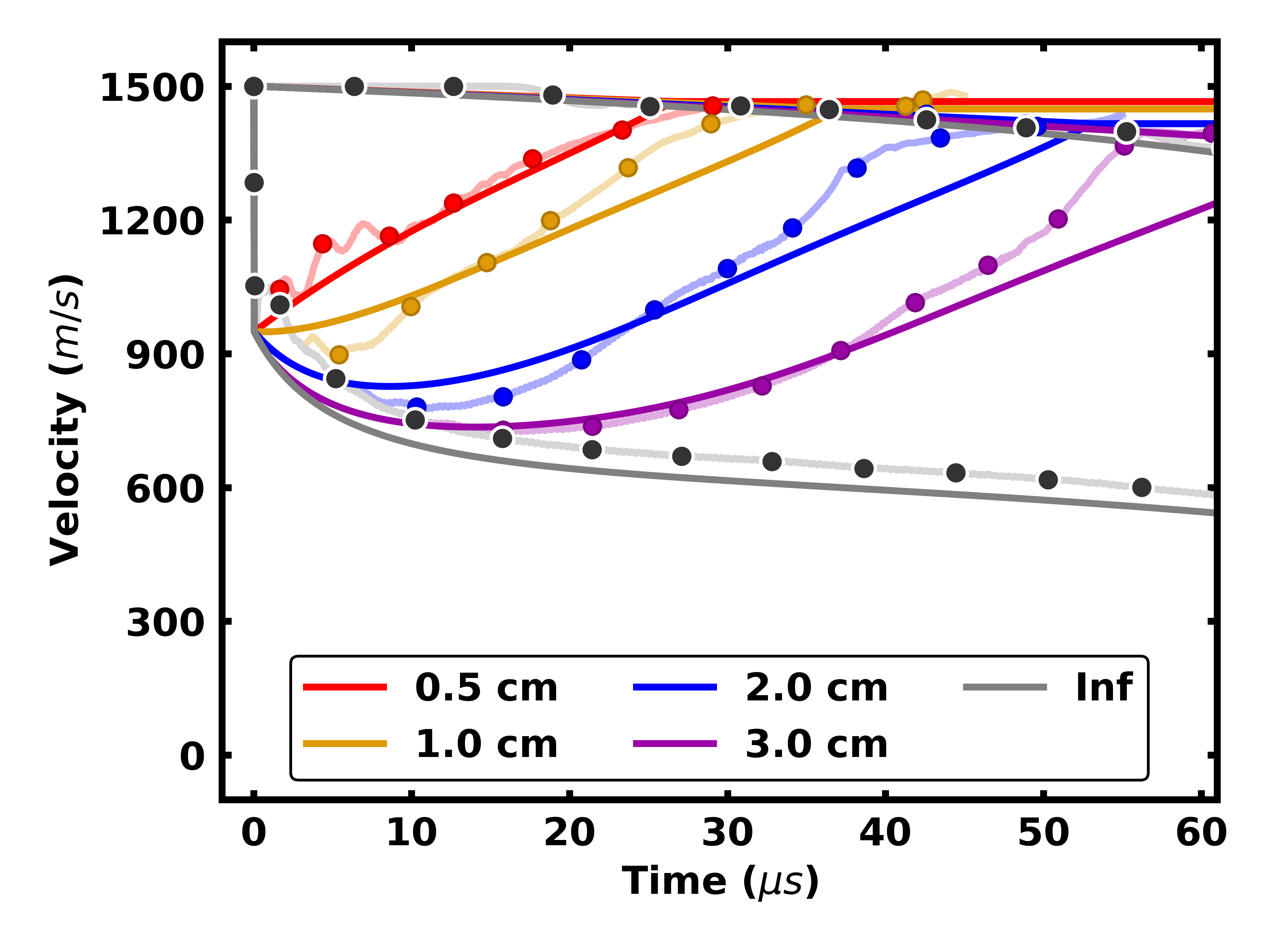}}
        \caption{Original Walker Back-face model vs ALE3D}
        \label{fig:walkerNoWave}
    \end{subfigure}
    \caption{Velocity time histories of nose and tail for impacts at 1500 m/s for 5 targets.  Light-line dark-dot series indicate a ALE3D reference, while the simpler single-color line indicates model performance.}
    \label{fig:results1500}
\end{figure}

We start here by examining similar conditions to those employed in the previous sections at 1500 m/s with results shown in Figure~\ref{fig:results1500}.
Figure~\ref{fig:walkerWave} demonstrates our improved model, which we can compare against Walker's original back-face model shown in Figure~\ref{fig:walkerNoWave}, with both figures' results for five targets, one infinite and the remaining four finite.
In matched colors for each nose-tail pair, each target has a solid line for the analytic model and a lighter line with circular markers plotting tracer data from ALE3D.
  
We see principally two features in our new model performance.
As verified earlier in Figure~\ref{fig:aleAgainstInf2} we see that our estimation of departure time from the infinite reference is well modeled across all shown finite targets.
This, combined with our improved starting conditions, well-explains the early-time improvement we hoped for when devising our solution, relative to the Walker back-face model.
The second, and more surprising feature is that improved early response, augmented by our continued modeling of the phase-delay, yields exceptional late time agreement as well.
This unexpected outcome speaks to the validity of the assumed mechanics we have introduced in this paper.

\begin{figure}[H]
    \centering
    \begin{subfigure}[t]{0.48\textwidth}
        \centering
    \fbox{\includegraphics[width=0.98\textwidth]{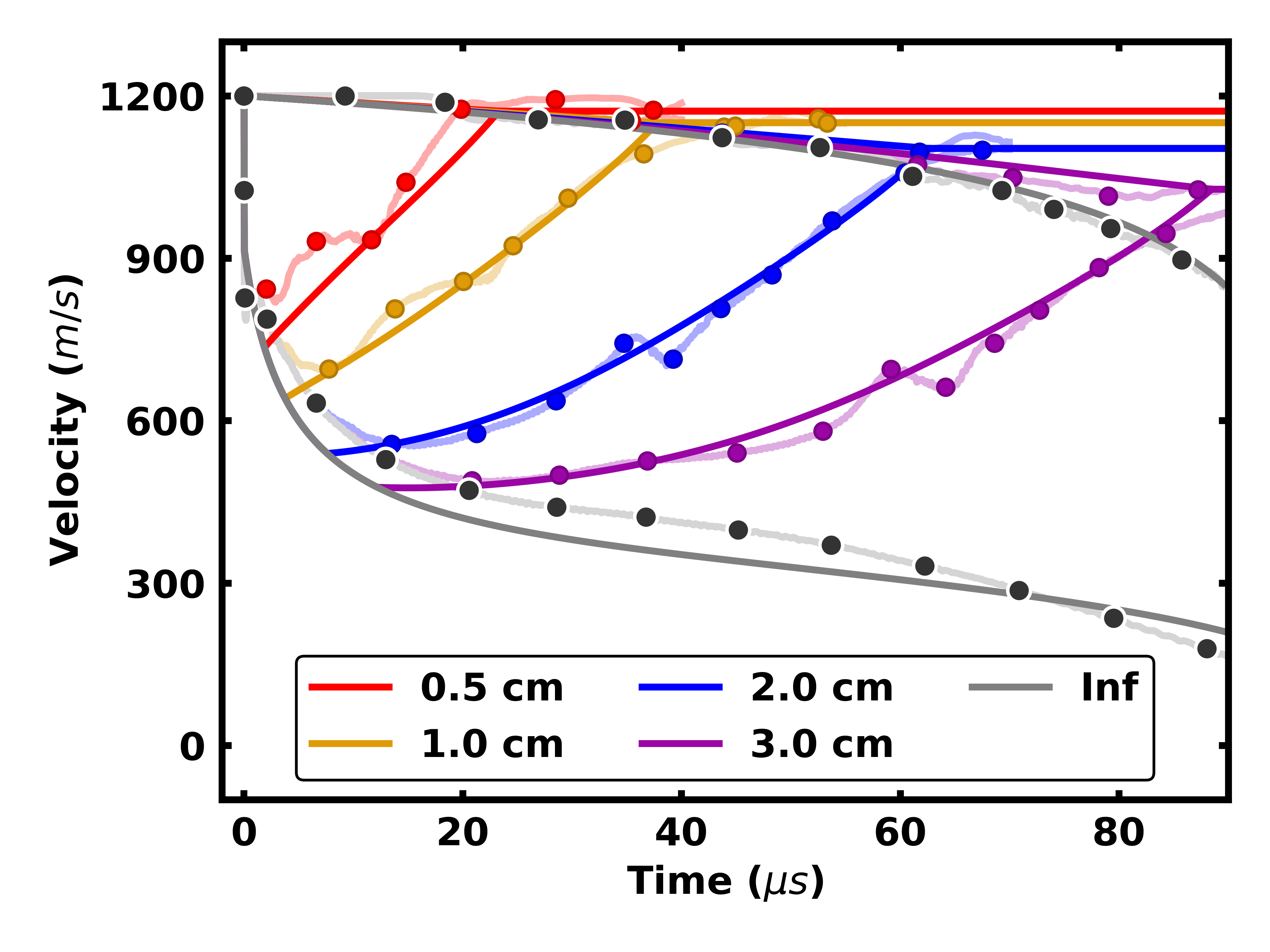}}
        \caption{1200 m/s}
        \label{fig:walker1200}
    \end{subfigure}
    ~
    \begin{subfigure}[t]{0.48\textwidth}
        \centering
    \fbox{\includegraphics[width=0.98\textwidth]{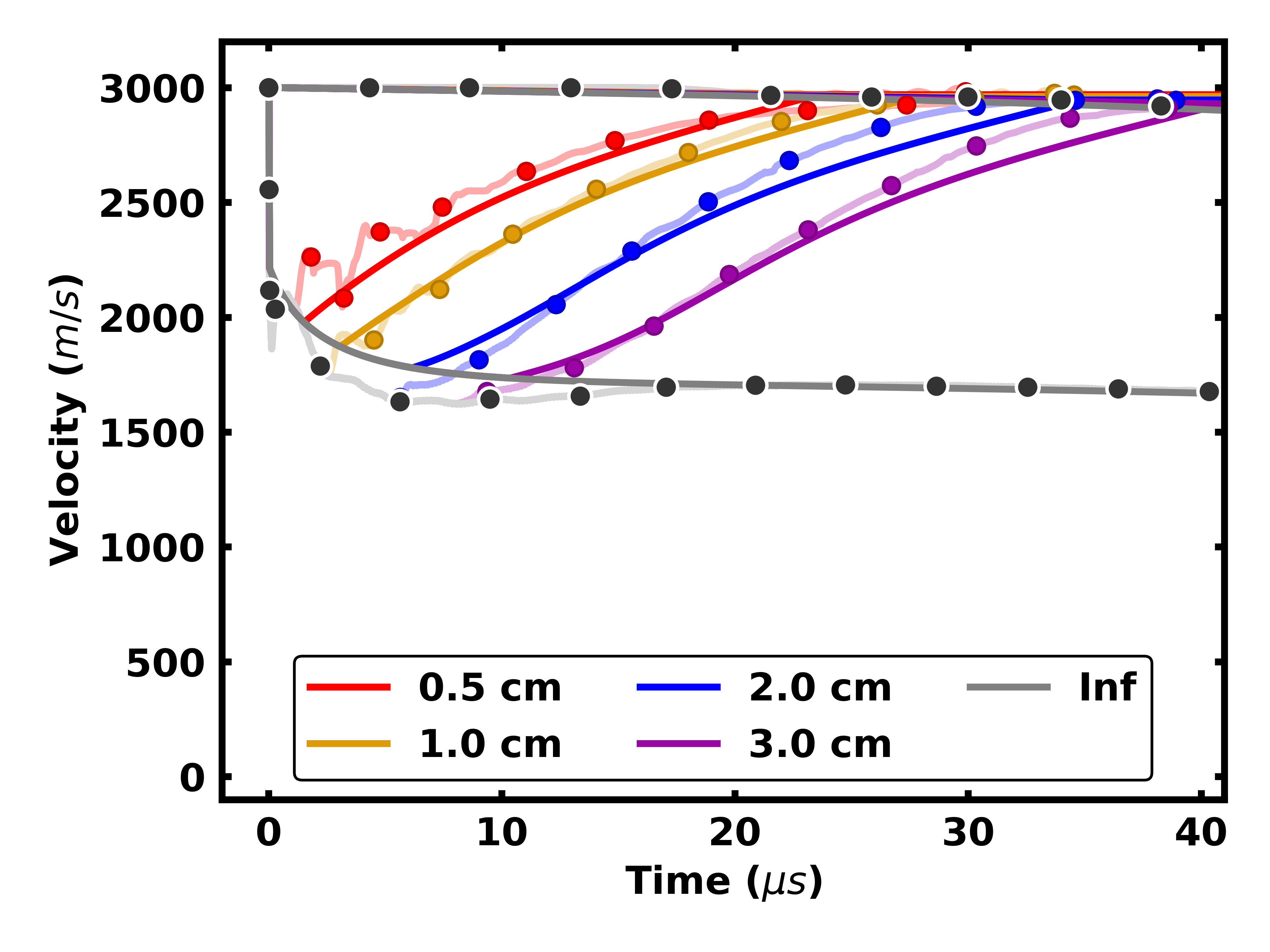}}
        \caption{3000 m/s}
        \label{fig:walker3000}
    \end{subfigure}
    \caption{Velocity histories for lower/higher impact velocity for 5 targets.  Light-line dark-dot series indicate a ALE3D reference, while the simpler single-color line indicates model performance. Note: vertical velocity axis is not matched in scale.}
    \label{fig:othervelplots}
\end{figure}

Having verified the model for conditions matching our work in deriving the model we now start by progressing to higher and lower velocities, shown in in Figure~\ref{fig:othervelplots}.
Note that in each plot, the time and velocity ranges do not match and vary as appropriate for the different initial conditions.

Figure~\ref{fig:walker1200} shows results for a slower 1200 m/s projectile, while Figure~\ref{fig:walker3000} has a faster 3000 m/s projectile.
Relative to our ALE3D references, our infinite target case of our 1200 m/s case predicts slightly lower nose velocities, whereas our 3000 m/s case predicts higher nose velocities.
However, the differences are not large given the complexity of the hydrodynamic problem, and the agreement for our finite targets through the recovery is excellent.
Given the quality of the results, we re-articulate that this degree of agreement arises naturally, and no parameters were re-tuned for these alternate speed impact velocities.

\begin{figure}[H]
    \centering
    \begin{subfigure}[t]{0.48\textwidth}
        \centering
    \fbox{\includegraphics[width=0.98\textwidth]{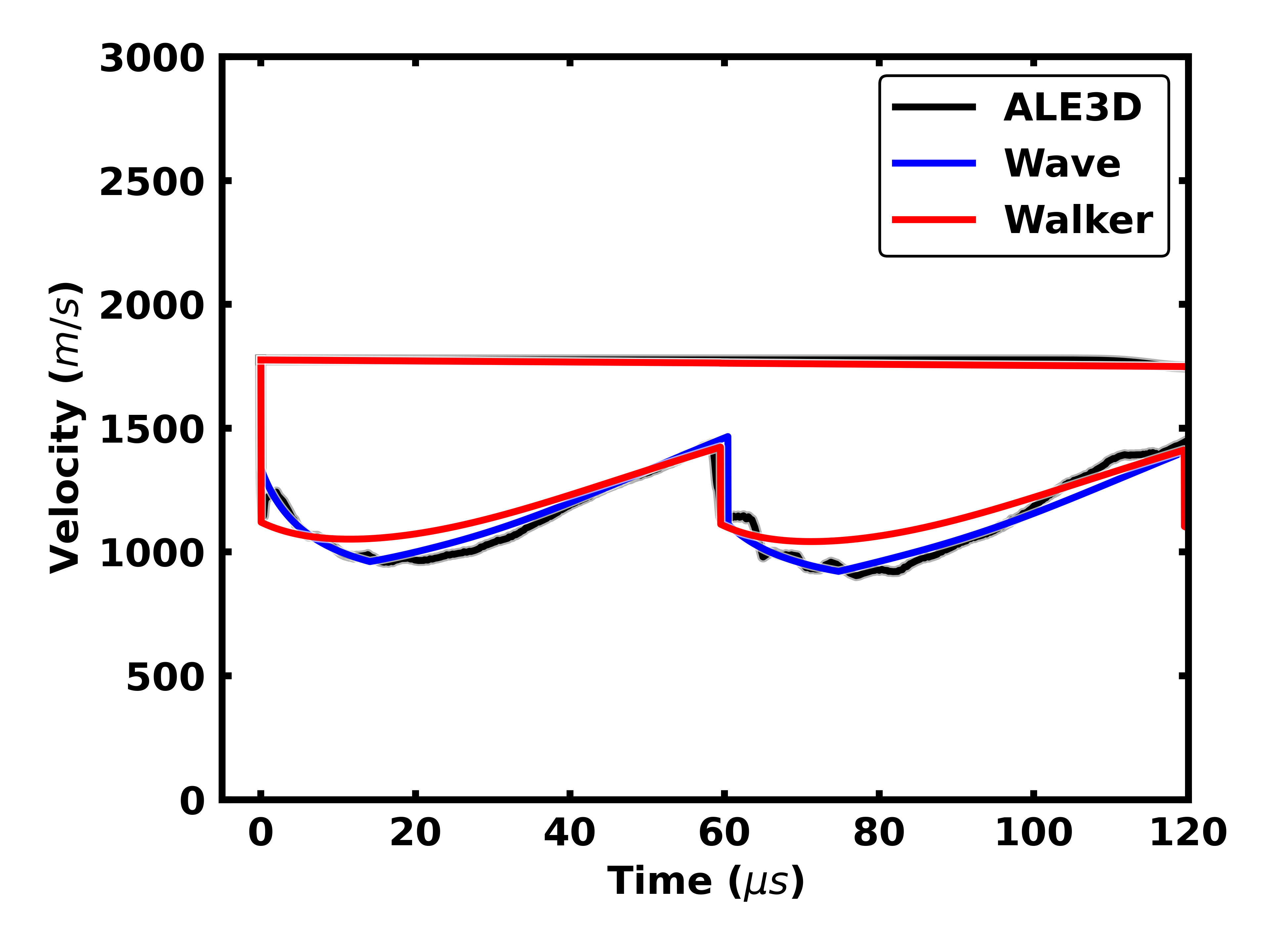}}
    \caption{First two plates of response}
        \label{fig:sixplateslowzm}
    \end{subfigure}
    ~
    \begin{subfigure}[t]{0.48\textwidth}
        \centering
    \fbox{\includegraphics[width=0.98\textwidth]{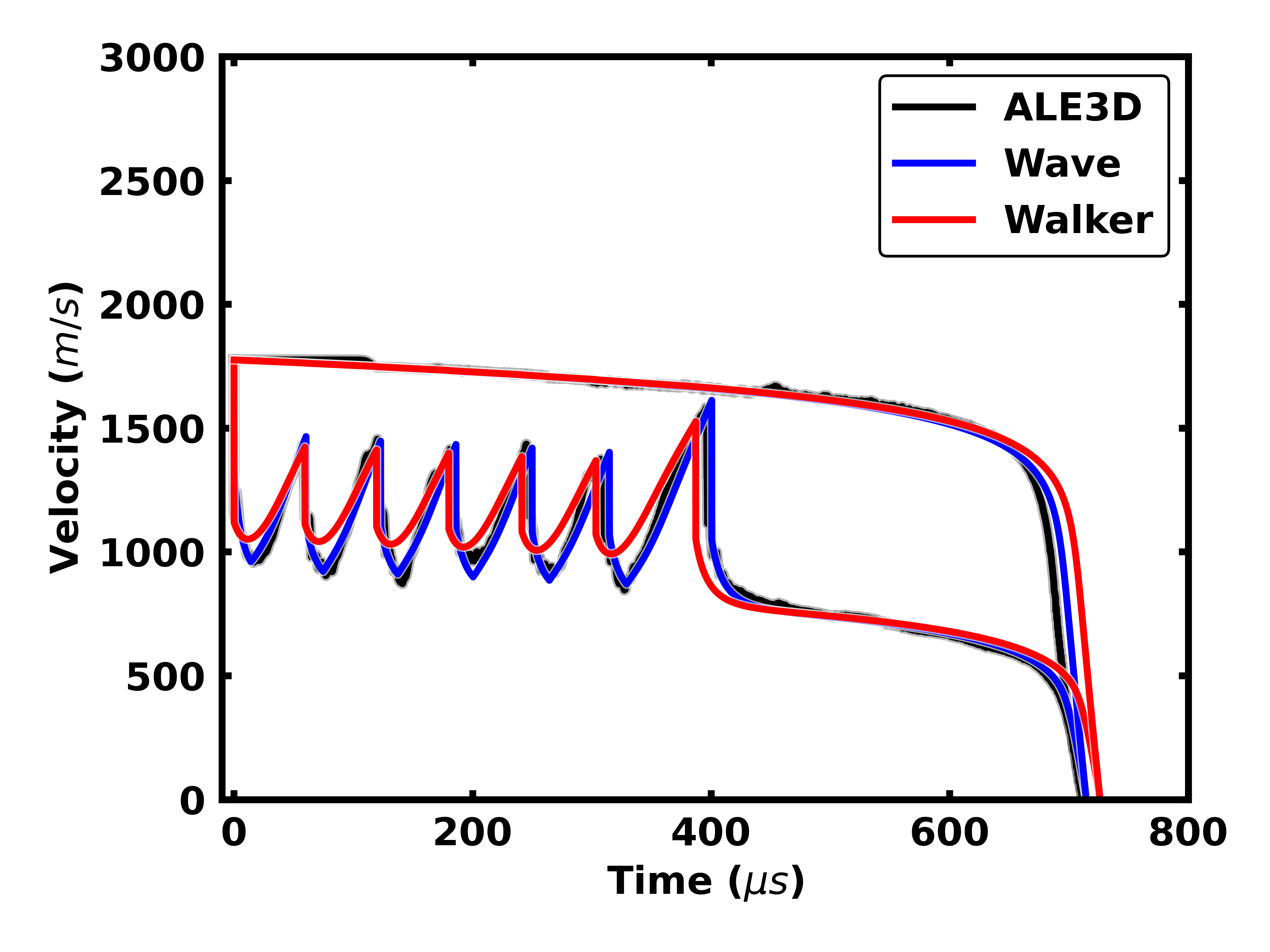}}
    \caption{Full simulation duration}
        \label{fig:sixplateslow}
    \end{subfigure}
    \caption{$V_0$=1775 m/s, Diam=1.65 cm, L=49.9 cm for six 4 cm targets and a witness plate}
    \label{fig:multitarget-slow}
\end{figure}

\begin{figure}[H]
    \centering
    \begin{subfigure}[t]{0.48\textwidth}
        \centering
    \fbox{\includegraphics[width=0.98\textwidth]{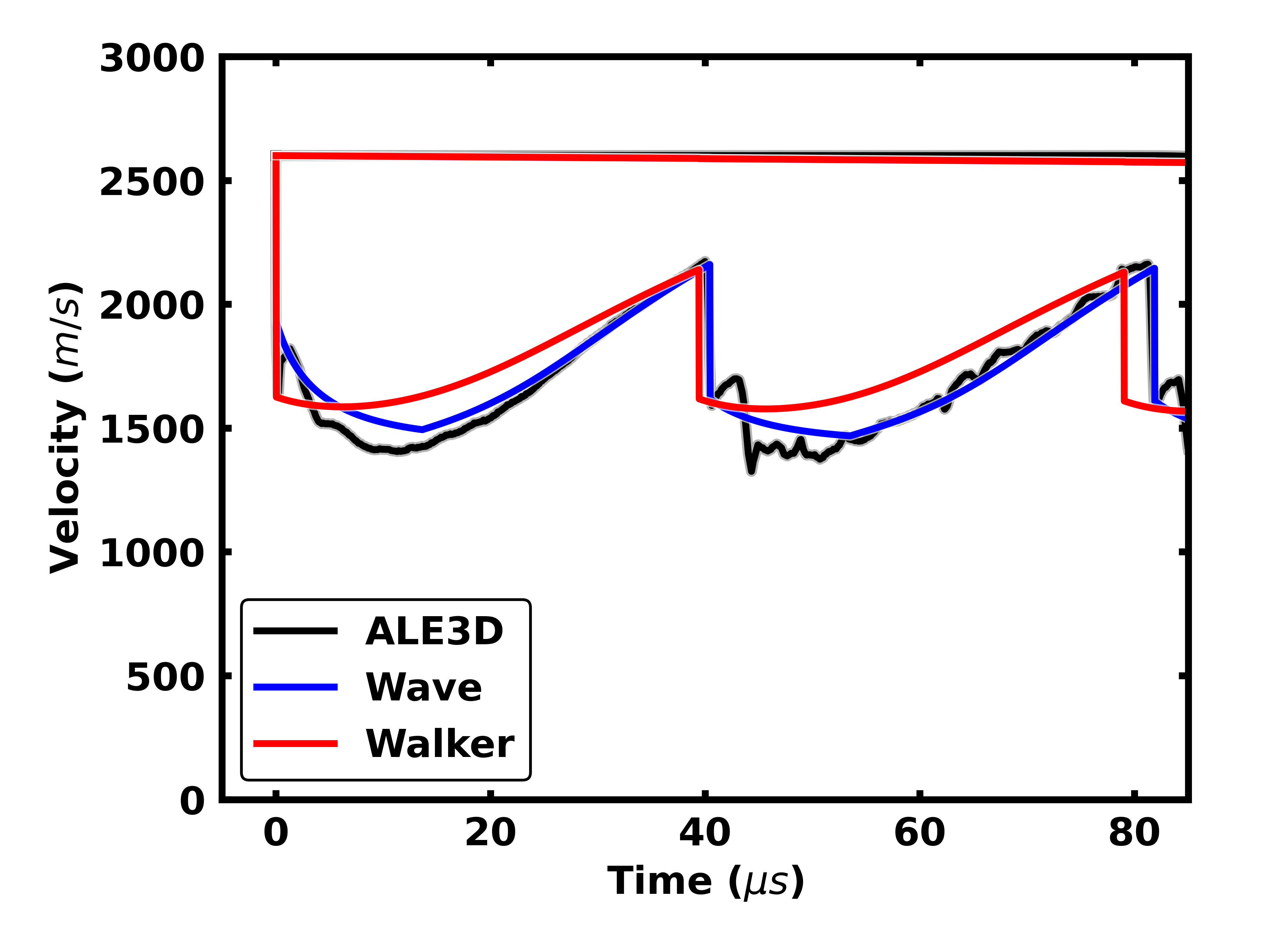}}
    \caption{First two plates of response}
        \label{fig:sixplatefastzm}
    \end{subfigure}
    ~
    \begin{subfigure}[t]{0.48\textwidth}
        \centering
    \fbox{\includegraphics[width=0.98\textwidth]{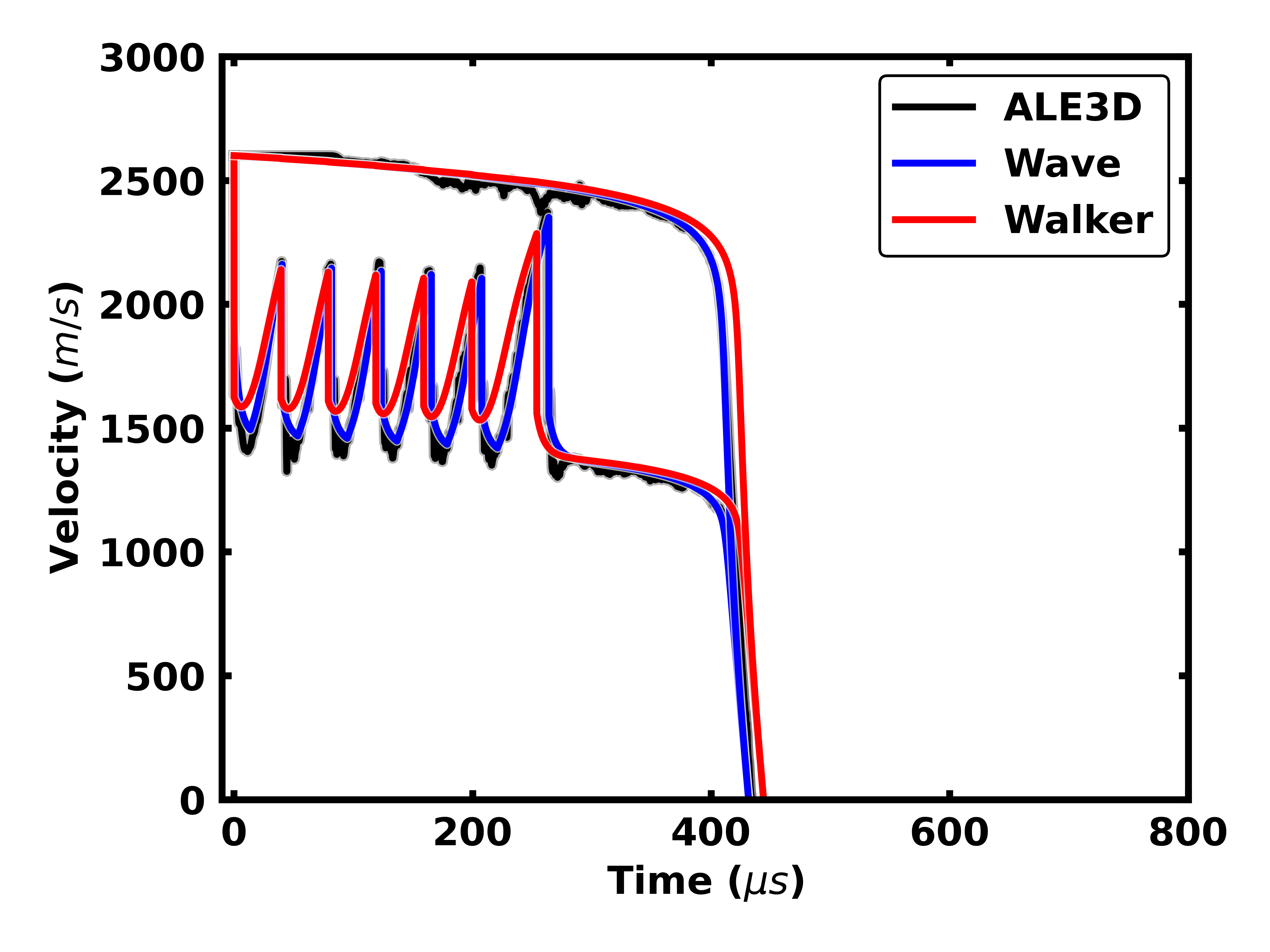}}
    \caption{Full simulation duration}
        \label{fig:sixplatefast}
    \end{subfigure}
    \caption{$V_0$=2600 m/s, Diam=1.28 cm, L=38.4 cm for six 4 cm targets and a witness plate}
    \label{fig:multitarget-fast}
\end{figure}

Having explored varying target thicknesses and impact speeds, we seek to further test the generality of our model.
For this purpose we borrow two multi-target examples \cite{Chocron2003} that articulate a significant change in system parameters.
The target set is comprised of six 4 cm thick plates gapped 3 cm apart followed by a half-space witness plate that is 6 cm behind the last plate.
Against this are two hemispherical-headed projectiles, each with its own velocity and diameter, with the dimensions and velocities listed in the Figure~\ref{fig:multitarget-slow} and \ref{fig:multitarget-fast} captions.
We start by examining our results for the slower case shown in Figure~\ref{fig:multitarget-slow} where the projectile is of similar velocity, but wider relative to the example used in our model development section.
Figure~\ref{fig:sixplateslowzm} focuses on the first two plates of the scenario, while \ref{fig:sixplateslow} allows us the see the general performance against the ALE3D reference over extended time.
What is apparent in both figures is that our new model is better able to capture initial response, as well as the sustained target strength, to the point that it seems unlikely this scenario could be better modeled analytically.

Moving to the faster, but not quite as wide, projectile shown in Figure~\ref{fig:multitarget-fast}, we see a similar set of short and long-time plots for our updated model.
Agreement here remains impressive, but we can see in Figure~\ref{fig:sixplatefastzm} that the somewhat high nose velocity estimates seen earlier for our 3000 m/s single-plate result are what keeps us from matching the depth of the troughs.
Despite that shortcoming, our inclusion of phase delay on target weakening allows the analytic target to induce a more sustained erosion of the impacting projectile relative to the Walker back-face model.

Ultimately, both considered multi-target scenarios represent an improvement of the updated model over the Walker's back-face model.
One of the impressive feats of Walker's back-face model extension was its ability to capture re-acceleration of the projectile nose without high sensitivity to a specific failure criteria.
As we look at it here though, we see that for thin targets the original model under-estimates target resistance.
Our introduction of a wave transit delay corrects this effect without any significant new parameters and only incurs a limited computational cost.

\section{Final Conclusions}

We set out in this work to improve semi-analytical modeling of finite targets by leveraging the accuracy of detailed ALE3D simulations.
By first studying hydrodynamic behaviors of the early-time impact response to half-space targets, we were able to define an improved definition of a starting condition for pristine hemi-headed projectiles.
The new effective-pressure condition better intializes the post-shock conditions for such penetration events.
With a new initialization of the projectile defined, we then set about improving the modeling of finite target effects.
By computing the velocity and plasticity wave propagation in a hydro-simulated finite target we numerically quantified a phase delay which is crucial to accounting for the effects of wave transit through a thin finite target.
Comparison against a thick-target test data further improved the model by validating the choice to omit a specific projectile recovery phase within the model.
Incorporating such effects allowed us to capture both the timing of transition to projectile recovery and nose recovery velocity profile in time.
The results demonstrate a marked improvement over the original Walker finite target penetration model, without the need for additional parameters or numerically intensive operations.
  
The agreement attained by these three corrections is such that it encourages a re-interpreting of the original Walker half-space model.
That model is centered around a steady state condition of penetration that is unconcerned about early time with no finite target effects.
Despite this, our fixes summarize to applying this half-space model both very early, immediately following impact, and also late, beyond the time when the original finite model would have started to bulge and weaken the target domain.
Fundamentally these applications are in conflict with the half-space model assumptions.
  
Of particular interest is that the tangential flow immediate to the projectile cavity more directly affects the nose-adjacent axial velocity.
Alternatively stated, the cavity's parting of material drives the extent of the plastic zone, but not in a way that immediately realizes the whole velocity field.
In this way we expect that velocities at the projectile nose would quickly match this field, but into the depths of the target we would then expect the corresponding field to develop with a delay.
In a steady erosion state this does not modify the original Walker understanding of the response, but it facilitates to realistically represent both early time and near-back-face conditions.
In early time, the conditions to flow around the hemisphere set up quickly, justifying the immediacy of the model's application.
In contrast to early time immediacy, this thinking then also validates a \textit{reduced} immediacy in terms of applying a finite back-face condition to the model.
The driving effect of the parting material then enhances a release wave that affects (e.g., interfere with the progression of) the cavity expansion.
In addition, this understanding of the flow near the plane tangent to the nose tip is itself consistent with the cylindrical cavity expansion used in the current and original model.
Though this choice was originally made based on empirical observations of test data, it now corroborates our new understanding of how flow dynamics drive target velocity-field propagation.

In essence the present model offers a remarkable accuracy in comparison to the high-fidelity hydrodynamic simulations that requires many times the computational resources and much more expertise to operate.
The proposed corrections to the existing semi-analytical methods are not extensive nor hard to implement in a fast-running algorithm.
The proposed corrections help the impact-engineering community better understand of coupled impactor-target interactions in high-rate penetration by an eroding penetrator.

\section{Acknowledgements}
This work was performed under the auspices of the U.S. Department of Energy by Lawrence Livermore National Laboratory under Contract DE-AC52-07NA27344 and was supported by the LLNL-LDRD Program under Project No. 25-SI-002. Lawrence Livermore National Security, LLNL-JRNL-2009603.
Additional thanks to Sidney Chorcon of the Southwest Research Institute for advise on paper improvements.
\bibliography{biblio} 
\bibliographystyle{elsarticle-num} 
\appendix

\section{Hydro-code material inputs} \label{sec:params}

In order to provide a reasonable baseline for the development of our model, we have need of material models that resolve the types of response we would expect in such regimes.
For this purpose, we here assemble tungsten and steel constitutive models from open literature parameters that will serve our purpose as reasonable representations that are not meant to match any particular experimental response.
The employed strength model is the classic Johnson-Cook model \cite{johnson1983constitutive} which captures rate hardening and thermal softening effects,
\begin{equation}
    \sigma_y = \left( A + B \varepsilon^n \right)\left( 1 + C \log{\dot{\varepsilon}^*} \right)\left( 1-(T^*)^m \right),
    \label{eq:jcstrength}
\end{equation}
where the specific parameters shown in Table \ref{tab:param-strength}.
\begin{table}[H]
\centering
\caption{Johnson-Cook Strength Parameters \vspace{-5pt}}
\label{tab:param-strength}
\begin{tabular}{rm{4em}m{4em}m{4em}m{4em}m{4em}m{4em}}  \hline \\[-8pt]
        & \centering $\mathrm{\boldsymbol{\rho}}$ \textbf{(g/$\mathbf{cm^3}$)}      & \centering \textbf{A (GPa)} & \centering \textbf{B (GPa)} & \textbf{n} & \textbf{C} & \textbf{m} \\ \hline \hline \\[-8pt]
    \textbf{Steel}   \cite{gray1994constitutive}        & \centering \phantom{-}7.85  & \centering 1.225            & \centering 1.575            & 0.768      & 0.0049     & 1.09       \\
      \textbf{Tungsten}  \cite{johnson1993material} & \centering 17.00           & \centering 1.506            & \centering 0.177            & 0.120      & 0.0160     & 1.00      \\ \hline
\end{tabular}
\end{table}

For the damage model we use the Johnson-Cook damage evolution\cite{JOHNSON1985}, 
\begin{equation}
    D = \sum \frac{\Delta \varepsilon_p}{\varepsilon^f}
    \label{eq:damage}
\end{equation}
where damage $D$ is updated in relation to the plastic increment $\Delta \varepsilon_p$, and estimated failure limit $\varepsilon^f$. 
This proceeds until unity is achieved at which time material will no longer carry shear strength.
The estimate failure form is,
\begin{equation}
    \varepsilon^f = \left( D_1 + D_2 \exp{D_3 \sigma^*} \right)\left( 1 + D_4 \log{\dot{\varepsilon}^*} \right)\left( 1 + D_5 T^* \right),
    \label{eq:failmodel}
\end{equation}
with the employed parameters and sources shown in Table \ref{tab:param-fail}.
\begin{table}[H]
\centering
\begin{threeparttable}[b]
\caption{Johnson-Cook Failure Parameters \vspace{-5pt}}
\label{tab:param-fail}
\begin{tabular}{rSSSSS}  \hline \\[-8pt]
    &   $\mathrm{\mathbf{D_1}}$   &  $\mathrm{\mathbf{D_2}}$ & $\mathrm{\mathbf{D_3}}$ & $\mathrm{\mathbf{D_4}}$ & $\mathrm{\mathbf{D_5}}$ \\ \hline \hline \\[-8pt]
    \textbf{Steel} \cite{buchar2002ballistic}     & -1.17\tnote{*}    & 2.10            & 0.50      & 0.002     & 0.61       \\
    \textbf{Tungsten} \cite{sun2020effect}        & 0.00              & 0.33            & 1.50      & 0.000     & 0.00      \\ \hline
\end{tabular}
\begin{tablenotes}
\item[*] \cite{buchar2002ballistic} uses 0.80
\end{tablenotes}
\end{threeparttable}
\end{table}
The only directly modified parameter is $D_1$ of the steel which was modified to give failure at 60\% strain in uniaxial tension.

What remains are the equations of state, which for both materials we take Mie-Gr\"{u}neisen parameters from Wilkins \cite{wilkins2013computer}.
\begin{table}[H]
\centering
\caption{Mie-Gr\"{u}neisen Parameters \vspace{-5pt}}
\label{tab:param-eos}
\begin{tabular}{rSSS}  \hline \\[-8pt]
    & $\mathrm{\mathbf{c_0\ (cm/\boldsymbol{\mu}s)}}$ & $\mathrm{\mathbf{k}}$    & $\mathrm{\mathbf{\gamma_0}}$    \\ \hline \hline \\[-8pt]
     \textbf{Steel}    \cite{wilkins2013computer}     &  0.46                    &  1.49     & 2.17                 \\
     \textbf{Tungsten} \cite{wilkins2013computer}     &  0.40                    &  1.24     & 1.54                 \\ \hline
\end{tabular}
\end{table}

\end{document}